\documentclass[fleqn,usenatbib,usedcolumn]{mnras}
   
\usepackage{mathptmx}
\usepackage{txfonts}

\usepackage[T1]{fontenc}
\usepackage{aecompl}

\usepackage{graphicx}	
\usepackage{subfig}

\def\multic#1#2{\multicolumn{#1}{c}{#2}}
\def\dagmark{\multic{1}{$\dag$\qquad}}

\hypersetup{pdfauthor={R. E. Ainsworth},
               pdftitle={GMRT detections of low mass young stars at 323 and 608 MHz},
               pdfkeywords={radio continuum: stars -- radiation mechanisms: thermal -- stars: pre-main-sequence -- stars: individual: L1551 IRS 5, T Tauri, DG Tauri},
               bookmarksnumbered=true}
   \volume{{\rm in press}}

\title[GMRT detections of YSOs]{GMRT detections of low mass young stars at 323 and 608\,MHz}

\author[R. E. Ainsworth et~al.]{
Rachael E. Ainsworth,$^{1}$\thanks{E-mail: rainsworth@cp.dias.ie}
Anna M. M. Scaife,$^{2}$
David A. Green,$^{3}$
Colm P. Coughlan,$^{1}$
\newauthor and Tom P. Ray$^{1}$
\\
$^{1}$Dublin Institute for Advanced Studies, School of Cosmic Physics, 31 Fitzwilliam Place, Dublin D02 XF86, Ireland\\
$^{2}$Jodrell Bank Centre for Astrophysics, School of Physics and Astronomy, The University of Manchester, Oxford Road, Manchester M13 9PL, UK\\
$^{3}$Astrophysics Group, Cavendish Laboratory, J J Thomson Avenue, Cambridge, CB3 0HE\\
}

\date{Accepted XXX. Received YYY; in original form ZZZ}

\pubyear{2015}

\begin{document}
\label{firstpage}
\pagerange{\pageref{firstpage}--\pageref{lastpage}}
\maketitle

\begin{abstract}
We present the results of a pathfinder project conducted with the Giant Metrewave Radio Telescope (GMRT) to investigate protostellar systems at low radio frequencies. The goal of these investigations is to locate the break in the free--free spectrum where the optical depth equals unity in order to constrain physical parameters of these systems, such as the mass of the ionised gas surrounding these young stars. We detect all three target sources, L1551~IRS~5 (Class~I), T~Tau and DG~Tau (Class~II), at frequencies 323 and 608\,MHz (wavelengths 90 and 50\,cm, respectively). These are the first detections of low mass young stellar objects (YSOs) at such low frequencies. We combine these new GMRT data with archival information to construct the spectral energy distributions for each system and find a continuation of the optically thin free--free spectra extrapolated from higher radio frequencies to 323\,MHz for each target. We use these results to place limits on the masses of the ionised gas and average electron densities associated with these young systems on scales of $\sim1000$\,au. Future observations with higher angular resolution at lower frequencies are required to constrain these physical parameters further.
\end{abstract}




\section{Introduction}
\label{intro}

The youngest protostars (Class~0 and I) are deeply embedded within embryonic envelopes and discs from which they accrete. They drive powerful bipolar outflows which likely remove angular momentum from the system and allow accretion to proceed. Class~II protostars, also known as classical T~Tauri stars, drive less powerful outflows as most of their mass has been accreted, but they are still surrounded by an optically thick disc \citep{2000prpl.conf...59A}. These outflows exhibit shocks and knots of many irregular morphologies along their lengths which can stretch to several parsecs and can terminate in bright bow shocks. The optical manifestations of these shocks are the so-called Herbig--Haro \citep[HH; see e.g.][]{2014arXiv1402.3553F} objects, and they have been detected in many bands of the electromagnetic spectrum from X-ray \citep[e.g.][]{2001Natur.413..708P} to radio wavelengths \citep[e.g.][]{2000AJ....119..882R}.

The radio emission from young stellar objects (YSOs) is typically detected as free--free radiation from the collimated, ionised outflows \citep[e.g.][]{2015aska.confE.121A}. The free--free spectrum is characterised by a flat or positive power-law spectral index $\alpha$ (where the flux density $S_{\nu}\propto\nu^{\alpha}$ at frequency $\nu$) and ranges between $-0.1$ for optically thin emission and $+2$ for optically thick emission. A value of $\alpha\approx0.6$ is the canonical value for a constant velocity, isothermal, spherical (or wide-angle) wind \citep[e.g.][]{1975A&A....39....1P} and values less than this (e.g. $\alpha\approx0.25$) are expected for more collimated flows \citep{1986ApJ...304..713R}. The HH objects typically exhibit optically thin free--free emission due to shock ionisation \citep[e.g.][]{1987RMxAA..14..595C, 1995RMxAC...1...59C, 2000AJ....119..882R}. 

The study of YSOs at radio wavelengths has been predominantly constrained to frequencies $>1$\,GHz ($\lambda<30$\,cm). This is due to the past sensitivity limitations of radio telescopes, the radio weakness of YSOs (flux densities $\la1$\,mJy at centimetre wavelengths) and the fact that their flux densities typically rise with frequency. However, recent surveys of nearby star forming regions with the Karl G. Jansky Very Large Array (VLA) find that at least half of the detected pre-main-sequence objects are non-thermal sources with active coronae characterised by high levels of variability, negative spectral indices, and in some cases significant circular polarisation \citep[e.g.][]{2013ApJ...775...63D, 2015ApJ...801...91D}. This is consistent with more evolved (Class~III) objects which have already shed most of their surrounding material and are detectable via the gyro-synchrotron emission from their coronae. 

Yet there are a small number of cases where non-thermal emission ($\alpha \ll -0.1$) is seen from Class~0--II YSOs. \citet{1997Natur.385..415R} detected circularly polarised gyro-synchrotron emission from the young embedded object T~Tauri~S, indicating that strong magnetic fields and at least mildly relativistic particles are present in the extended outflow of this low mass YSO. Furthermore, \citet{2010Sci...330.1209C} detected linearly polarised emission from the high mass protostellar jet, HH~80-81, which allowed for the direct measurement of the magnetic field direction and estimates of its strength in this outflow. These results are remarkable, as the presence of relativistic particles was previously unexpected from these outflows which have typical shock velocities of order $\sim30-300$\,km\,s$^{-1}$. More importantly, they shed light on the large-scale magnetic field which has remained an elusive characteristic of these systems. Finally, there are a handful of low mass Class~0--II YSOs with outflow components which exhibit spectral indices indicative of non-thermal emission \citep[e.g.][]{1993ApJ...415..191C, 2002RMxAA..38..169G}, however linearly polarised emission has yet to be detected in the outflow from a low mass YSO.

We conducted a pathfinder project with the Giant Metrewave Radio Telescope (GMRT) to search for YSOs at very low radio frequencies ($<1$\,GHz). We observed a sample of three well-studied low mass YSOs with known radio emission previously measured at frequencies $>1$\,GHz: L1551~IRS~5, T~Tau and DG~Tau. All three reside in the Taurus Molecular Cloud at distances of approximately $130-140$\,pc \citep{1994AJ....108.1872K, 2012ApJ...747...18T}. The GMRT has a spatial resolution of approximately 10 and 5\,arcsec at 325 and 610\,MHz, respectively. We are therefore exploring these systems on large scales of $\sim700-1400$\,au.

The importance of investigating the radio emission from young stars at low radio frequencies is twofold. First, observations of the low frequency turnover in the free--free spectrum (at frequency $\nu_0$) can constrain physical properties of the ionised plasma associated with these systems. Where the free--free emitting plasma is reasonably uniform in density, there is a characteristic radio spectrum which has two components, optically thick behaviour at lower frequencies ($\tau_\nu\gg1$, with $S_\nu\propto\nu^{2}$) and optically thin behaviour at higher frequencies ($\tau_\nu\ll1$, with $S_\nu\propto\nu^{-0.1}$). They are delineated by the frequency at which the optical depth equals unity ($\tau_{\nu_0}=1$) which is dependent on the size, density and temperature of the ionised plasma \citep[see e.g.][]{2013AdAst2013E..14S}. The first two of these quantities are combined into the emission measure of the plasma ($EM=\int n_{\rm e}^2 {\rm d}s$, where $n_{\rm e}$ is the electron density and integration is conducted along the distance $s$). The turnover frequency $\nu_0$ is required to break the degeneracy between source size and gas mass inherent within the characteristic property of emission measure in these objects. Once the physical size of the emitting region is constrained, the total mass of the ionised gas ($M_{\rm ion}$) associated with the YSO can be determined. 

Low frequency observations of young stars are also important as the flux densities of non-thermal emission processes increase with decreasing frequency which should make them easier to detect. From the current set of observations, emission with a synchrotron spectral index was detected in the proximity of the jet driven by the target source DG~Tau which was interpreted as arising from a prominent bow shock associated with this outflow. This result, which provides tentative evidence for the acceleration of particles to relativistic energies due to the shock impact of this otherwise very low-power jet against the ambient medium, was presented in \citet{2014ApJ...792L..18A}.

In this paper we present the lowest frequency detections of YSOs made to date. In Section~\ref{sec:observations} we present details of the observations and data reduction. In Section~\ref{sec:results} we present radio images made at 323 and 608\,MHz (90 and 50\,cm, respectively) of the three target YSOs with the GMRT, measure their associated flux densities and calculate their spectral indices. In Section~\ref{sec:discussion} we discuss the detections and non-detections, and model the spectral energy distributions (SEDs) for each target source. We find that these new GMRT detections show a continuation of the free--free spectrum extrapolated from higher frequencies, and that the frequency at which the emission transitions between optically thin and thick behaviour has not yet been reached. We therefore place limits on the mass of ionised gas, the average electron density and the emission measure, assuming the turn-over frequency is less than 323\,MHz. We make our concluding remarks in Section~\ref{sec:conclusions}.

\section{Observations and Data Reduction}
\label{sec:observations}

The observations were made using the GMRT \citep[see e.g.][]{2005ICRC...10..125A}, which consists of thirty $45$\,m dishes. Fourteen of the thirty dishes are arranged in a compact, quasi randomly distributed central array within a square kilometre. The remaining sixteen dishes are spread out along three arms of an approximately ``Y''-shaped configuration, with a longest interferometric baseline of about 25\,km. 

Observations of the young stars L1551~IRS~5, T~Tau and DG~Tau were made at 325 and 610\,MHz between 6--14 December 2012 (average epoch 2012.95). At 325\,MHz, observations were taken for 7\,hours per night over the course of three nights for a total of 21\,hours and observations at 610\,MHz were taken for 10\,hours in a single run. See Table~\ref{tab:sourcelist} for the number of hours on-source for each target. A total bandwidth of 32\,MHz was observed, which was split into 256 spectral channels. The sample integration time was 16.9\,s. The primary beam of the GMRT has a full width at half-maximum (FWHM) of approximately 85\,arcmin at 325\,MHz and 44\,arcmin at 610\,MHz. The observational details can be found in Table~\ref{tab:sourcelist}.

\begin{table*}
\centering
\caption{Observing details. Column [1] contains the target source name; [2] the protostellar evolutionary class, [3] the source Right Ascension; [4] the source Declination; [5] the observing frequency (pre-processing, see Section~\ref{sec:observations}); [6] the observing wavelength; [7] the on-source observing time; [8] the dimensions of the full width at half-maximum of the synthesised beam; and [9] the rms noise (see Section~\ref{sec:results} for details).}
\label{tab:sourcelist}
\begin{tabular}{lcccccccc} 
\hline
Source & Class & \multicolumn{2}{c}{J$2000.0$ Coordinates} & $\nu$ & $\lambda$ & Obs. time & FWHM, PA & $\sigma_{\rm{rms}}$  \\
       & & $\alpha$ ($^{\rm h}~~^{\rm m}~~^{\rm s}$) & $\delta$ ($^{\circ}$~$'$~$''$) & (MHz) & (cm) & (hrs.) & ($''\times''$, $^\circ$) & ($\umu$Jy\,beam$^{-1}$)  \\
\hline
L1551~IRS~5 & I & 04~31~34.1 & +18~08~04.8 & $325$ & $90$ & $6.0$ & $11.4\times9.5$, $-88.5$ & $151$  \\
	& & & & $610$ & $50$ & $2.2$ & $6.2\times4.9$, $76.5$ & $49$ \\
T~Tau & II & 04~21~59.4 & +19~32~06.4 & $325$ & $90$ & $3.3$ & $10.8\times9.5$, $-81.6$ & $103$ \\
	& & & & $610$ & $50$ & $2.2$ & $6.0\times5.0$, $83.8$ & $45$ \\
DG~Tau & II & 04~27~04.7 & +26~06~16.3 & $325$ & $90$ & $6.0$ & $11.6\times9.2$, $79.6$ & $127$ \\
	& & & & $610$ & $50$ & $2.2$ & $6.5\times5.2$, $74.0$ & $80$ \\
\hline
\end{tabular}
\end{table*}

The flux density scale was set through observations of 3C48 at the beginning and end of each observing run. For the cases where 3C48 set before the end of the run, 3C147 or 3C286 were observed. The flux densities were calculated using the Astronomical Image Processing Software (\textsc{AIPS}) task \textsc{setjy} \citep{2013ApJS..204...19P} and were found to be 45.6\,Jy at 325\,MHz and 29.4\,Jy at 610\,MHz for 3C48, 55.1\,Jy at 325\,MHz for 3C147, and 20.8\,Jy at 610\,MHz for 3C286. Each target source was observed for a series of interleaved 10\,min scans so as to maximise the \textit{uv}~coverage for imaging. The nearby phase calibrator, J0431+206, was observed for 3\,min after every two scans so as to monitor the phase and amplitude fluctuations of the telescope. The AIPS task \textsc{getjy} retrieved flux densities of $2.78\pm0.02$\,Jy at 325\,MHz and $3.05\pm0.05$\,Jy at 610\,MHz for J0431+206.
 
Flagging of baselines, antennas, channels, and scans that suffered heavily from interference was performed for each night's observation using standard AIPS tasks. Bandpass calibration was applied to each antenna using the flux calibrator sources. Ten central frequency channels were then combined together with the task \textsc{splat} and antenna-based phase and amplitude calibration was performed with \textsc{calib}. This calibration was then applied to the full dataset and averaged into 24 separate spectral channels. Some end channels, where the bandpass correction is larger, were omitted. The effective mean frequency of the observations was therefore 322.665 and 607.667\,MHz (hereafter referred to as 323 and 608\,MHz, respectively throughout this paper) with an effective bandwidth of 30\,MHz covered by the averaged 24 channels. Target source data was then extracted from the larger datasets via the \textsc{split} task and the observations from each run were concatenated with \textsc{dbcon} for imaging. 

The large field of view of the GMRT leads to significant phase errors if the whole field is imaged directly, due to the non-planar nature of the sky \citep[see e.g.][]{2007MNRAS.376.1251G}. To minimise these errors, we followed \citet{2007MNRAS.376.1251G} and conducted wide-field imaging using facets. Each field was divided into 61 smaller facets for the 323\,MHz images and into 31 facets for the 608\,MHz images. The facets were imaged separately, each with a different assumed phase centre, and then recombined into a hexagonal grid. In each case the total area covered by the facets is larger than the FWHM of the GMRT primary beam which allows bright sources well outside of the observed region to be cleaned from the images. This technique was used to better clean the full GMRT field of view to achieve the best possible sensitivity at the phase centre as well as to obtain information over the entire field in order to create a catalogue of sources. The task \textsc{setfc} was used to create a list of facet positions for use in \textsc{imagr}. Images were made with a pixel size of 3\,arcsec at 323\,MHz and 1.5\,arcsec at 608\,MHz to adequately sample the synthesised beams. 

Each field went through three iterations of phase self-calibration using a model dominated by the bright sources in the field at 10, 3 and 1\,min intervals, and then a final round of self-calibration correcting both phase and amplitude errors at 10\,min intervals \citep[following][]{2007MNRAS.376.1251G}. The overall amplitude gain was held constant so that the flux density of sources was unaffected. These self-calibration steps improved the noise levels up to 30\,per~cent and significantly reduced the residual side lobes around bright sources. Images of each field were produced with \textsc{robust} set to 0 within \textsc{imagr} to optimise the trade-off between angular resolution and sensitivity. The resulting facets were then stitched together with \textsc{flatn} and corrected for the primary beam using an eighth-order polynomial with coefficients taken from the GMRT Observer's Manual\footnote{http://www.ncra.tifr.res.in/ncra/gmrt/gmrt-users/observing-help-for-gmrt-users/gmrt-observers-manual-07-july-2015/view} within \textsc{pbcor}. 

A visual inspection of the overall target fields suggests that a systematic displacement might be present. A number of bright, compact (presumably extragalactic) sources in each field were fit using \textsc{jmfit} in AIPS at both 323 and 608\,MHz to compare the absolute positions. In the majority of cases, the positions at 323\,MHz differ from those at 608\,MHz by approximately 0.86\,arcsec in Declination for the L1551~IRS~5 field, 0.17\,arcsec for the T~Tau field and 2\,arcsec for the DG~Tau field. The shift is understandable due to the difference in the ionosphere with Declination between the target and phase calibrator (up to $\approx6^\circ$ separation for DG~Tau), which should be worse at 323\,MHz than 608\,MHz. The positions at 608\,MHz are in general consistent with those from the NRAO VLA Sky Survey \citep[NVSS,][]{1998AJ....115.1693C} to within an average of 0.5\,arcsec, making them more reliable than the positions at 323\,MHz. However, due to a lack of low frequency radio surveys at higher resolution for this patch of the sky (NVSS has a spatial resolution of approximately 45\,arcsec) the absolute coordinates may have a residual uncertainty of around 0.5\,arcsec. This should not have a significant impact as the GMRT spatial resolution is of order 5--10\,arcsec. We therefore shifted the 323\,MHz images by 0.86\,arcsec to the south in Declination for the L1551~IRS~5 field and 2\,arcsec to the south in Declination for the DG~Tau field to make them more consistent with the 608\,MHz images. The 323\,MHz image in the case of the T~Tau field has not been shifted as this target was closest in position to the phase calibrator and does not suffer heavily from this problem.

A full catalogue of sources detected within the FWHM primary beam of each target field, including a detailed description of the survey methodology and data products, will be presented in a forthcoming paper (Ainsworth et~al., in prep).

\section{Results}
\label{sec:results}

\begin{figure*}
\subfloat{\includegraphics[width=0.43\textwidth]{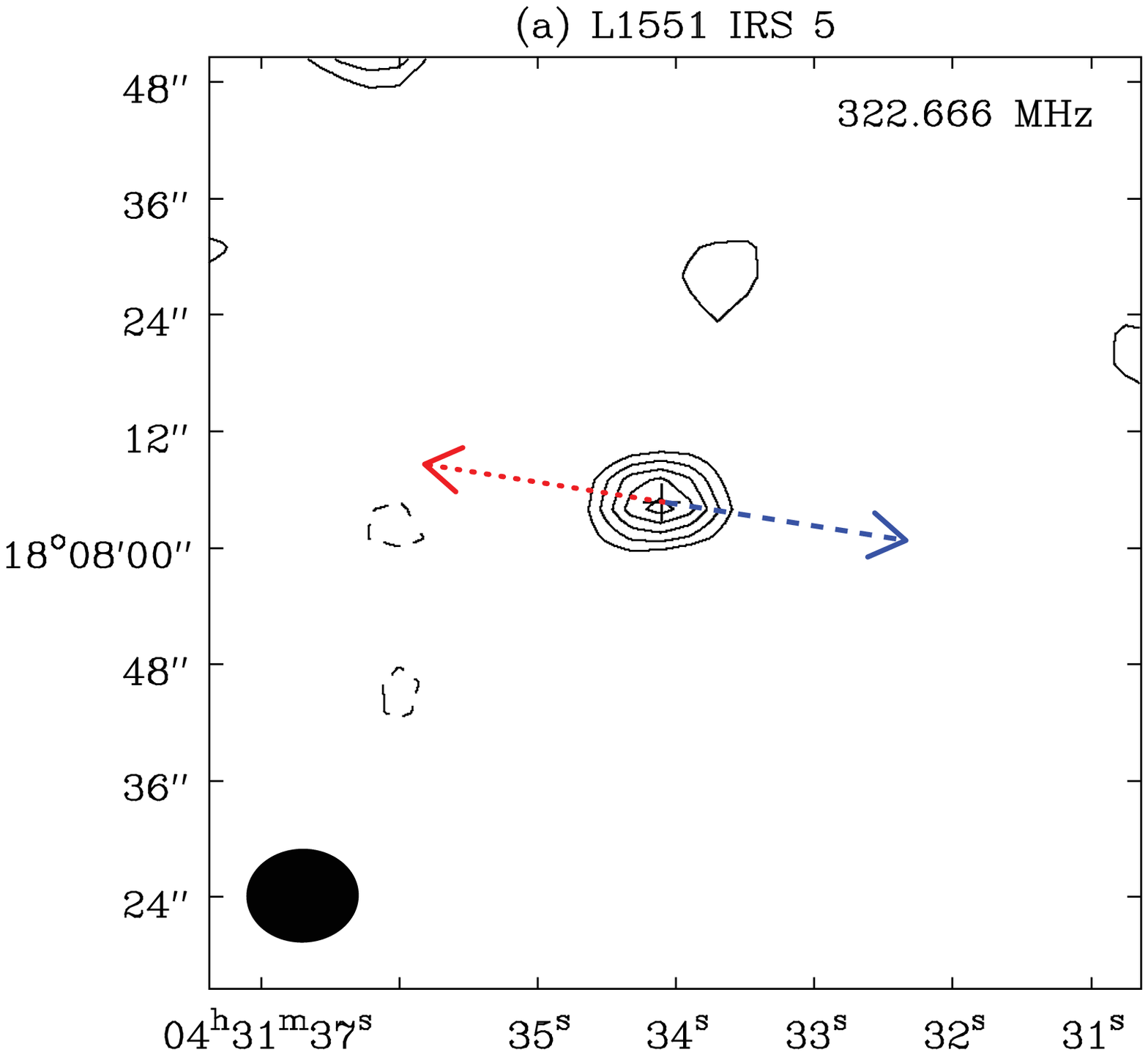} \label{L1551_323}}
\subfloat{\includegraphics[width=0.43\textwidth]{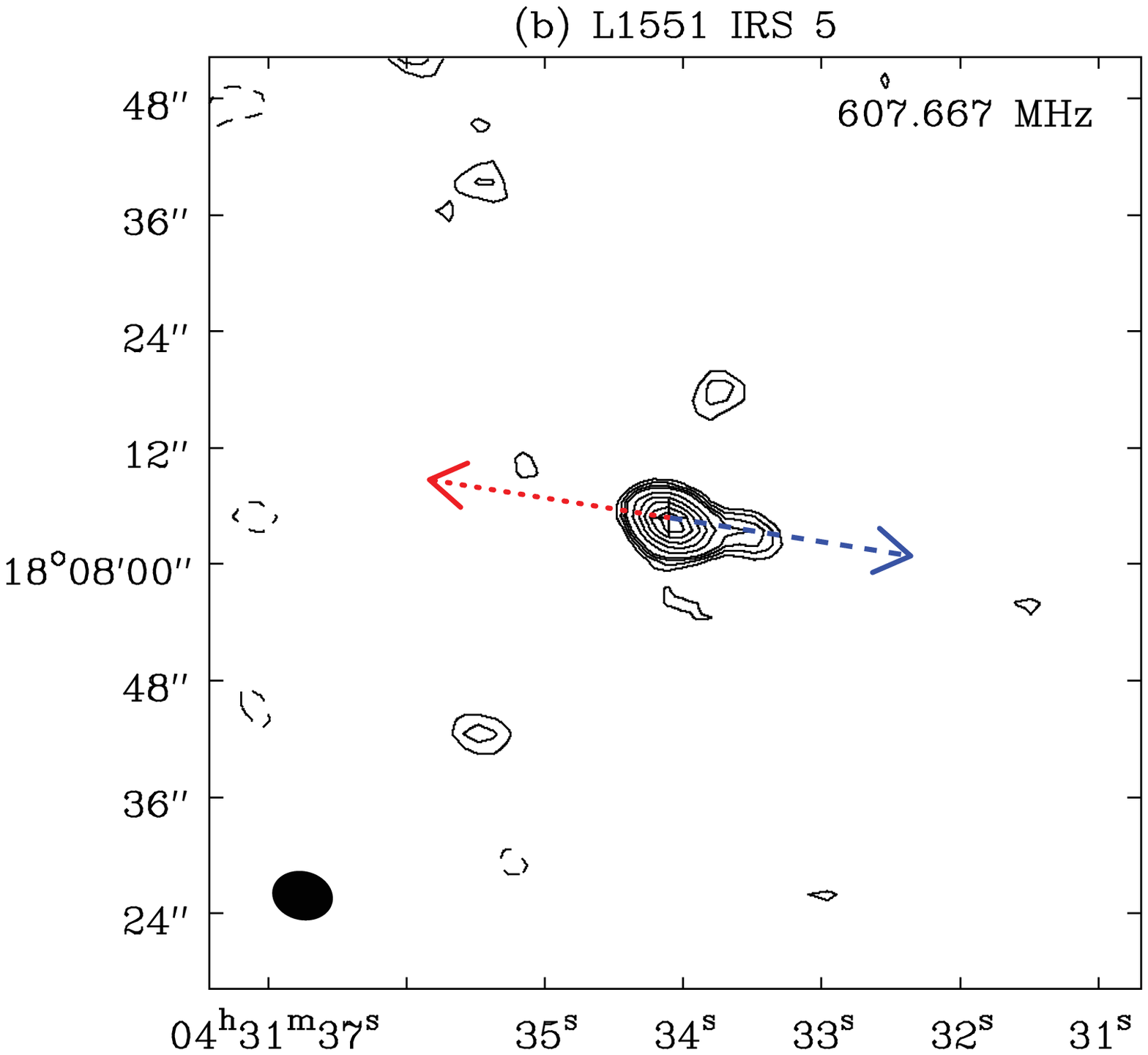} \label{L1551_608}} \\
\subfloat{\includegraphics[width=0.43\textwidth]{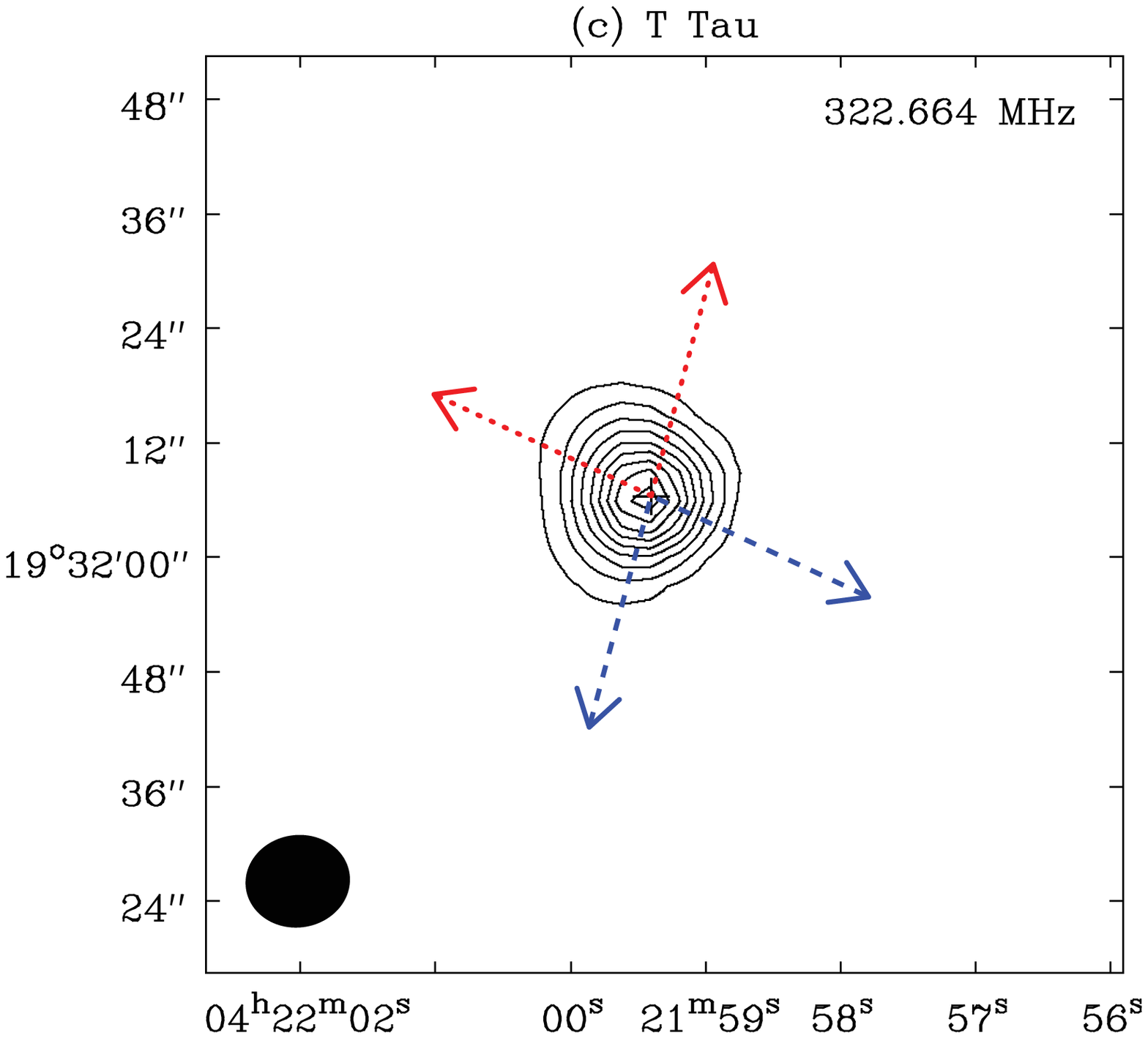} \label{TTAU_323}}
\subfloat{\includegraphics[width=0.43\textwidth]{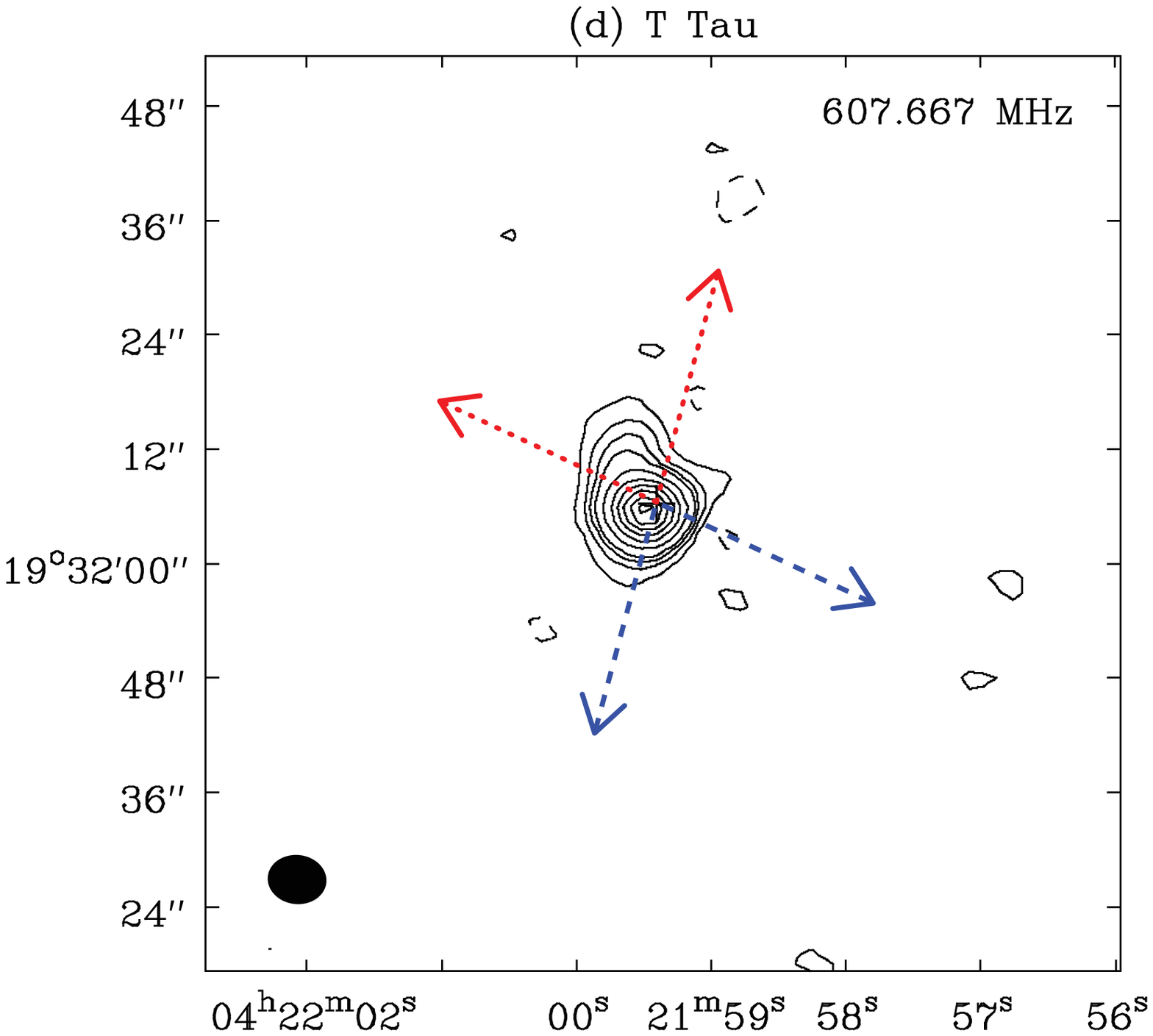} \label{TTAU_608}} \\
\subfloat{\includegraphics[width=0.43\textwidth]{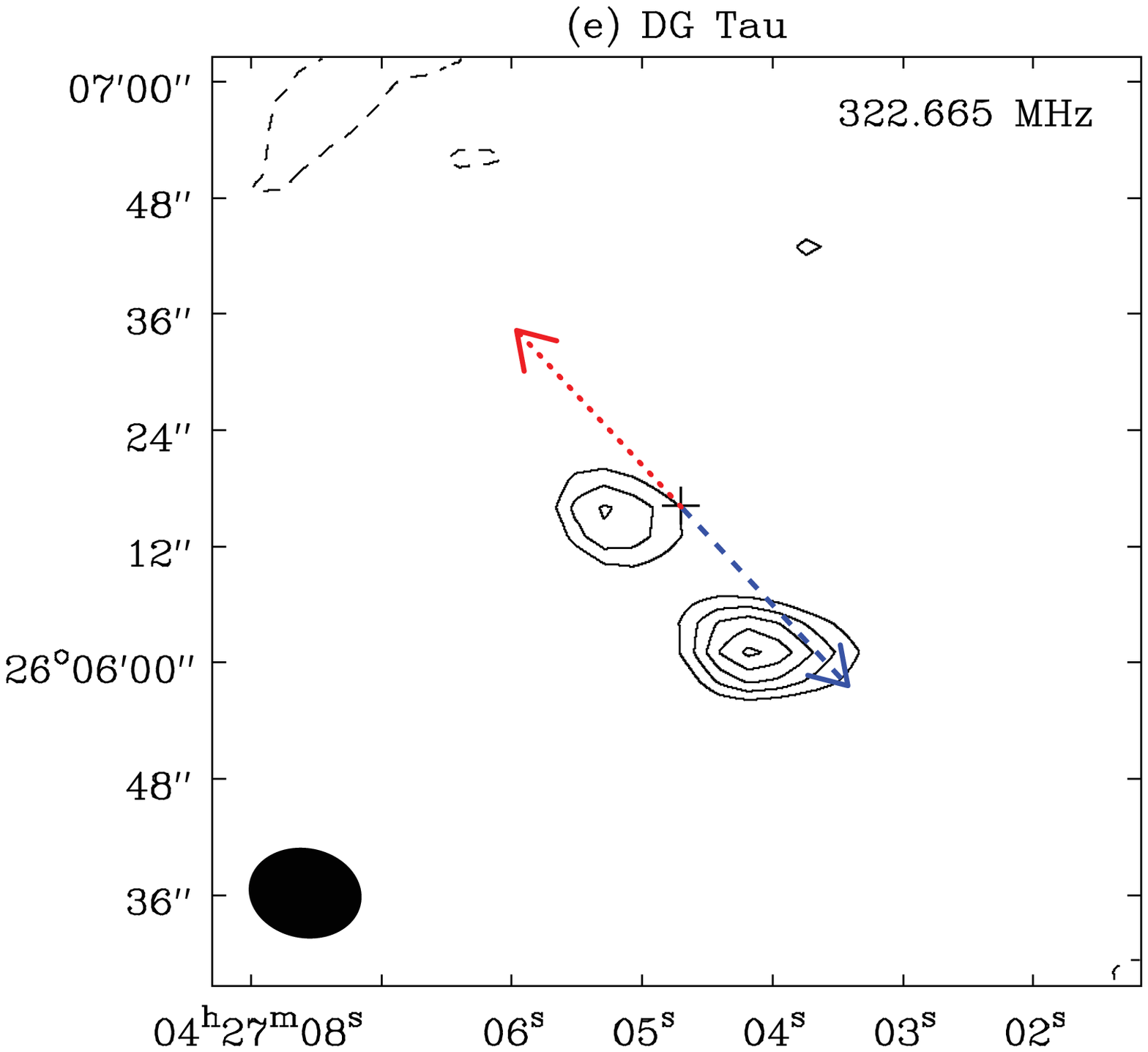} \label{DGTAU_323}}
\subfloat{\includegraphics[width=0.43\textwidth]{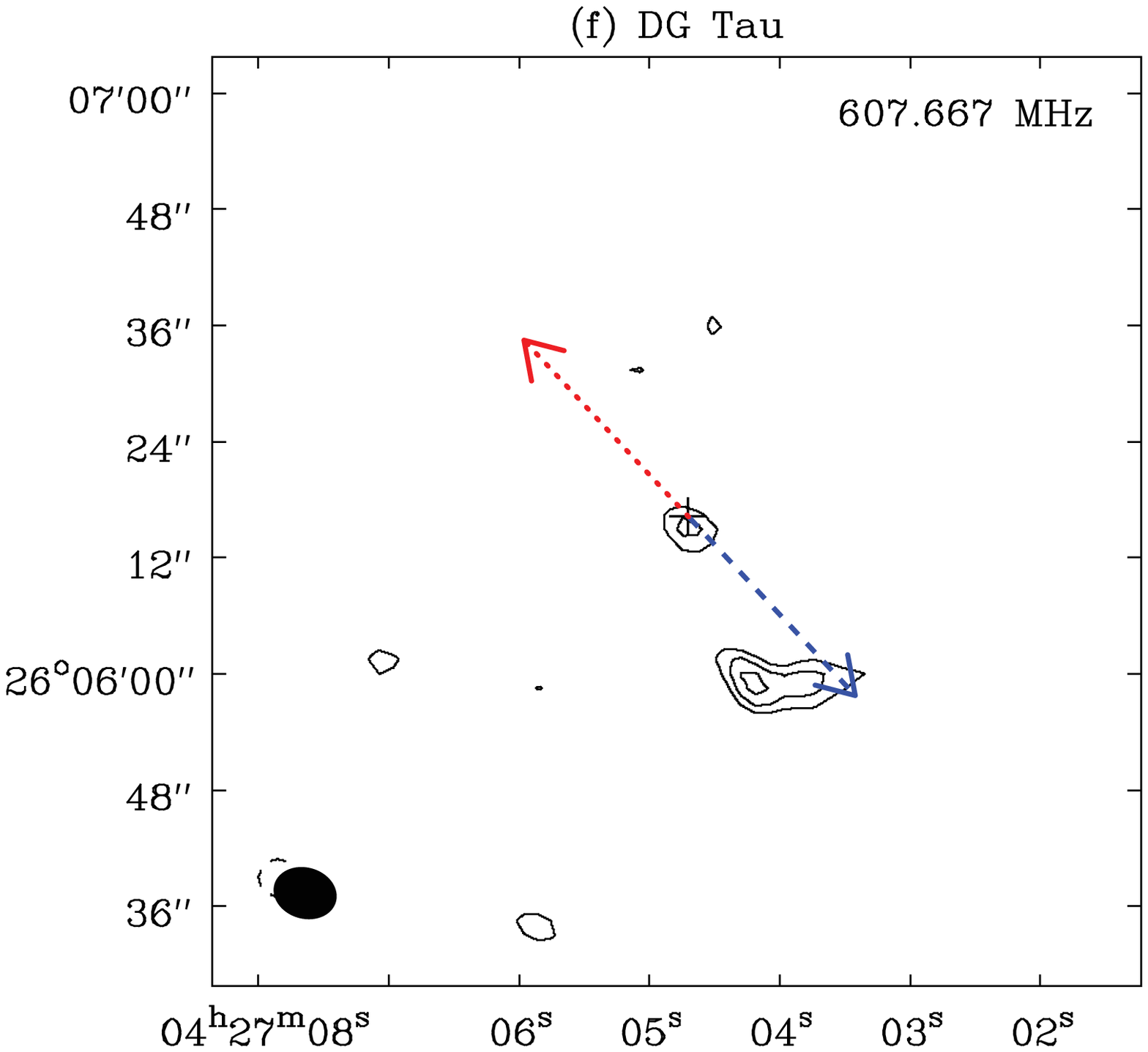} \label{DGTAU_608}}
\caption{GMRT combined channel images for (a) L1551~IRS~5 at 323\,MHz with contour levels $-3, 3, 4, 5, 6, 7 \times \sigma_{\rm rms}$, (b) L1551~IRS~5 at 608\,MHz with contour levels $-3, 3, 4, 5, 6, 9, 12, 15, 18, 21 \times \sigma_{\rm rms}$, (c) T~Tau at 323\,MHz with contour levels $-3, 3, 6, 9, 12, 15, 18, 21, 24, 27 \times \sigma_{\rm rms}$, (d) T~Tau at 608\,MHz with contour levels $-3, 3, 6, 9, 12, 20, 30, 40, 50, 60, 70 \times \sigma_{\rm rms}$, (e) DG~Tau at 323\,MHz with contour levels $-3, 3, 4, 5, 6, 7 \times \sigma_{\rm rms}$ and (f) DG~Tau at 608\,MHz with contour levels $-3, 3, 4, 5 \times \sigma_{\rm rms}$. The values for $\sigma_{\rm rms}$ for each source are listed in Table~\ref{tab:sourcelist}. The coordinates are J2000.0, the synthesised beam is shown as a filled ellipse in the bottom left corner of each image and the stellar positions are denoted by $+$ symbols. Approximate, large-scale red and blue-shifted outflow axes are denoted as dotted (red) and dashed (blue) lines, respectively (see Sections~\ref{sec:L1551}, \ref{sec:TTAU} and \ref{sec:DGTAU} for details). A colour version of this figure is available in the online journal.}
\label{fig:maps}
\end{figure*}

Combined channel maps centred at 323 and 608\,MHz for each target source are presented in Fig.~\ref{fig:maps}. Target sources were identified with a peak flux density $>3\sigma_{\rm rms}$, where $\sigma_{\rm rms}$ is the root-mean-square (rms) noise in a local patch of sky using the AIPS task {\sc imean} and listed in Table~\ref{tab:sourcelist}. Flux densities were extracted with {\sc jmfit} through Gaussian fitting. Errors on the flux densities are calculated as $\sigma_{S_{\nu}} = \sqrt{(0.05S_{\nu})^2+\sigma_{\rm{fit}}^2}$, where $0.05S_{\nu}$ is an estimated 5\,per~cent absolute calibration error on the flux density $S_{\nu}$ and $\sigma_{\rm fit}$ is the fitting error returned from {\sc jmfit}. In this case, {\sc jmfit} estimates the error in the fits using the image rms determined from the image header keyword {\sc actnoise}, the values of which were determined using {\sc imean} and are listed in Table~\ref{tab:sourcelist}. Flux densities measured with {\sc jmfit}, their associated errors, and deconvolved component dimensions are listed in Table~\ref{tab:fluxes}. All errors quoted are $1\sigma_{S_\nu}$.

\begin{table*}
\centering
\caption{{\sc jmfit} results. Column [1] contains the target source name; [2] the peak flux density at 323\,MHz; [3] the integrated flux density at 323\,MHz; [4] the source deconvolved dimensions at 323\,MHz; [5] the peak flux density at 608\,MHz; [6] the integrated flux density at 608\,MHz; [7] the source deconvolved dimensions at 608\,MHz; and [8] the spectral index extrapolated between the two GMRT frequencies for each source as calculated in Section~\ref{sec:alphas}. Flux densities, associated errors and deconvolved dimensions were measured using the \textsc{AIPS} task \textsc{jmfit}, see Section~\ref{sec:results} for details.}
\label{tab:fluxes}
\begin{tabular}{lccccccc} 
\hline
Source & $S_{\rm peak, 323\,MHz}$ & $S_{\rm int, 323\,MHz}$ & $\theta_{\rm 323\,MHz}$ & $S_{\rm peak, 608\,MHz}$ & $S_{\rm int, 608\,MHz}$ & $\theta_{\rm 608\,MHz}$ & $\alpha_{\rm GMRT}$ \\
       & (mJy\,beam$^{-1}$) & (mJy) & ($''\times''$, $^\circ$) & (mJy\,beam$^{-1}$) & (mJy) & ($''\times''$, $^\circ$) &  \\
\hline
L1551~IRS~5 & $1.03\pm0.16$ & $1.38\pm0.32$ & $10.4\times0.0$, 119.0 & $1.15\pm0.08$ & $1.40\pm0.12$ & $4.6\times0.0$, 68.1 & $0.02\pm0.39$ \\
T~Tau & $2.94\pm0.18$ & $3.95\pm0.29$ & $7.5\times3.7$, 7.4 & $3.22\pm0.17$ & $4.24\pm0.23$ & $3.6\times2.2$, 6.0 & $0.11\pm0.14$ \\
DG~Tau & $<0.38$ &  &  & $0.33\pm0.08$ & $0.58\pm0.20$ & $7.0\times3.6$, 70.0 & \\ 
\hline
\end{tabular}
\end{table*}

\subsection{Spectral Indices}
\label{sec:alphas}

The spectral index between the GMRT frequencies observed was calculated for each source using
\begin{equation}
\alpha_{\rm GMRT}=\frac{\ln(S_{\rm int, 323\,MHz}/S_{\rm int, 608\,MHz})}{\ln({323/608)}}
\end{equation}
where $S_{\rm int, 323\,MHz}$ and $S_{\rm int, 608\,MHz}$ are the integrated flux densities at frequencies 323 and 608\,MHz, respectively (see Table~\ref{tab:fluxes}). The error on the spectral index, $\sigma_{\alpha_{\rm GMRT}}$, was calculated using standard propagation of error theory \citep[e.g.][]{Ku1966}, where
\begin{equation}
\sigma_{\alpha_{\rm GMRT}}=\frac{1}{|\ln(323/608)|} \left[ \left( \frac{\sigma_{S_{\rm int, 323\,MHz}}}{S_{\rm int, 323\,MHz}}\right)^{2} + \left(\frac{\sigma_{S_{\rm int, 608\,MHz}}}{S_{\rm int, 608\,MHz}} \right)^{2} \right]^{1/2}
\end{equation}
and $\sigma_{S_{\rm int, 323\,MHz}}$ and $\sigma_{S_{\rm int, 608\,MHz}}$ are the errors on the integrated flux densities at 323 and 608\,MHz, respectively (see Table~\ref{tab:fluxes}). The spectral indices between the two GMRT frequencies for each source and their associated errors are listed in Table~\ref{tab:fluxes}.

\section{Discussion}
\label{sec:discussion}

\subsection{Notes on Individual Targets}

In this section, we limit our discussion to the three YSO targets of these GMRT observations and their results. The Taurus Molecular Cloud is not as densely populated with young stars as other star forming regions \citep[e.g.][]{2015ApJ...801...91D}, however there are several other known pre-main-sequence objects located within the GMRT field of view (primary beam FWHM is approximately 85\,arcmin at 323\,MHz and 44\,arcmin at 608\,MHz) for each of our target fields. For example, the entirety of the Lynds~1551 (L1551) star forming region (within the larger Taurus Molecular Cloud complex) fits within the GMRT field of view at 608\,MHz. Therefore, the GMRT can be a potentially useful survey instrument for star forming regions at very long wavelengths due to the extent of its field of view. A full survey comprising of all three fields, including a detailed description of the survey methodology, data products and additional YSO detections, will be presented in a forthcoming paper.

\subsubsection{L1551~IRS~5}
\label{sec:L1551}

The target L1551~IRS~5 is a Class~I, deeply embedded and optically invisible multiple protostellar system that was discovered by \citet{1976AJ.....81..320S} in a near-infrared survey of the L1551 star forming region. It was the first object from which a molecular outflow was discovered \citep{1980ApJ...239L..17S} and more recent observations in CO reveal a complex array of molecular outflow activity \citep{2006ApJ...649..280S, 2009ApJ...698..184W}. It is a binary system consisting of Northern and Southern components \citep{1985ApJ...289L...5B} which are separated by $\sim0.3$\,arcsec ($\sim42$\,au at 140\,pc). 

With high angular resolution at radio wavelengths, each component is shown to be comprised of a circumstellar dust disc and bipolar ionised jets \citep{1998Natur.395..355R, 2003ApJ...586L.137R}. The high-velocity jets are aligned with the large-scale outflow which has a PA of $261^\circ$ \citep{2009ApJ...694..654P}.  \citet{2006ApJ...653..425L} report a third component approximately 13\,au southeast of the Northern component as another circumstellar dust disc, suggesting that IRS~5 is a triple system. The IRS~5 sources are nested within a larger $\sim10\,000$\,au envelope \citep{2002A&A...382..573F}.

The GMRT does not resolve the separate sources, but detects the overall integrated emission of the IRS~5 system. At 323\,MHz, the L1551~IRS~5 system is a point source with respect to the synthesised beam (see Fig.~\ref{L1551_323}). At 608\,MHz however, the emission appears resolved and extended to the west (see Fig.~\ref{L1551_608}) consistent with the large-scale outflow direction. The GMRT spectral index $\alpha_{\rm GMRT}=0.02\pm0.39$ is in general agreement with the overall radio spectral index $\alpha$ determined using measurements at shorter wavelengths (see Section~\ref{sec:SEDs}) which is consistent with optically thin free--free emission.

\subsubsection{T~Tau}
\label{sec:TTAU}

Historically, T~Tau has served as the prototype for an entire class of YSOs \citep{1945ApJ...102..168J}, but has proved to be increasingly interesting over recent years. It is a Class~II triple system consisting of T~Tau~N, an optically visible star, and T~Tau~S, a heavily extincted binary system (T~Tau~Sa and Sb) approximately 0.7\,arcsec south of T~Tau~N \citep{1982ApJ...255L.103D,2000ApJ...531L.147K}. The southern binary is completely undetectable at visible wavelengths \citep{1998ApJ...508..736S}, but is brighter than T~Tau~N at $\lambda>3\,\umu$m and may dominate the overall energetics \citep{1991AJ....102.2066G,1997AJ....114..744H}. 

There are two prominent outflows associated with the T~Tau system \citep[see e.g.][and references therein for an overview of the system]{2007ApJ...657..916L}. One is oriented approximately east--west (PA $\sim65^\circ$) which terminates at HH~155 in the west and is driven by T~Tau~N. The second outflow is oriented approximately southeast--northwest (PA $\sim345^\circ$), is associated with the HH objects 255 and 355, and has a total projected extent of 1.55\,pc \citep{1997AJ....114.2708R}. T~Tau~Sa is believed to power the large-scale southeast--northwest outflow, whilst T~Tau~Sb drives a separate molecular outflow in the southwest direction \citep{2010A&A...517A..19G}. 

The T~Tau outflow is also the only low mass Class~II YSO outflow for which circularly polarised emission has been detected \citep[e.g.][]{1994AJ....107.1461S, 1997Natur.385..415R, 2003AJ....125..858J},  indicating that at least some of the emission is magnetic in origin. The very high resolution observations carried out by \citet{1997Natur.385..415R} with MERLIN at 5\,GHz whilst the source was undergoing an outburst suggest the presence of at least mildly relativistic electrons spiralling along strong, ordered magnetic field lines in the outflow. Spectral indices have been reported for T~Tau~S ranging from values consistent with partially optically thick free--free emission \citep[e.g. $\alpha=0.44\pm0.2$,][]{1986ApJ...303..233S} to slightly negative or optically thin free--free \citep[e.g. $\alpha=-0.2\pm0.2$,][]{1994AJ....107.1461S}. 

The radio emission at 323 and 608\,MHz for T~Tau with the GMRT is resolved with respect to the synthesised beam and shows a clear extension to the north (see Fig.~\ref{TTAU_323} and \ref{TTAU_608}). The GMRT does not resolve the separate YSOs but detects the overall emission of the T~Tau system. This morphology is consistent with previous radio investigations \citep[e.g.][]{1982ApJ...253..707C, 1986ApJ...303..233S, 2003AJ....125..858J} as both T~Tau~N and S produce radio emission, although the radio emission here is likely dominated by T~Tau~S. The morphology is also consistent with the ``A'' bubble of H$_{2}$ emission described by \citet{1997AJ....114..744H}. The spectral index between the GMRT data is $\alpha_{\rm GMRT}=0.11\pm0.14$. It is in agreement with the overall radio spectral index $\alpha$ determined using measurements at shorter wavelengths (see Section~\ref{sec:SEDs}) which is indicative of optically thin free--free radiation.

\subsubsection{DG~Tau}
\label{sec:DGTAU}

DG~Tau is a highly active Class~II object driving a bipolar outflow, although the blueshifted jet is better studied. It was one of the first T~Tauri stars to be associated with an optical jet and exhibits several shocks and knots in its outflow \citep[HH~158,][]{1983ApJ...274L..83M}, the most prominent of which are currently located about 5 and 12\,arcsec from the star. The optical blueshifted jet has a PA of $223\degr$ close to the source \citep{1997A&A...327..671L}, however large-scale studies trace the jet out to a total projected distance of approximately 0.5\,pc and show that the PA of the outflow changes to $218\degr$ \citep[HH~702,][]{2004A&A...420..975M}. Radial velocities in the jet have been found to range up to 350\,km\,s$^{-1}$, with average velocities of 200\,km\,s$^{-1}$ \citep{2000A&A...357L..61D}. The small-scale redshifted optical (counter) jet is difficult to see owing to absorption/extinction by the foreground extended disc structure \citep{1996ApJ...457..277K} but exhibits morphological and kinematic asymmetries with the blueshifted flow \citep[e.g.][]{1997A&A...327..671L, 2014MNRAS.442...28W}. There is no known large-scale redshifted HH outflow associated with DG Tau. 

Extended soft ($<1$\,keV) X-ray emission has been detected from the DG~Tau jet out to 5\,arcsec coincident with the optical outflow, and both hard ($>1$\,keV) and soft emission has been observed from the source \citep{2008A&A...478..797G}. In the radio, DG~Tau has been shown to have a compact and elongated morphology close to the source in the direction of the optical outflow and possesses a spectral index typical of free--free emission at frequencies between 5--20\,GHz \citep{1982ApJ...253..707C, 1986AJ.....92.1396C, 2012AA...537A.123R, 2012MNRAS.420.3334S, 2013ApJ...766...53L, 2013MNRAS.436L..64A}. 

At 323 and 608\,MHz, DG~Tau is detected at 5 and $4\sigma_{\rm rms}$, respectively, and appears point-like (see Fig.~\ref{DGTAU_323} and \ref{DGTAU_608}). Furthermore, there is a significant signal detected to the southwest of DG~Tau which is extended in an approximate east--west direction. The investigations of this emission to the southwest were presented in \citet{2014ApJ...792L..18A} which identified it as the radio counterpart to an optical bow-shock associated with this outflow \citep{1998AJ....115.1554E} and showed that it exhibits a synchrotron spectrum. The non-thermal nature was attributed to particle acceleration in the shock of the outflow against the denser ambient medium.

There is, however, a clear offset between the radio emission at the two GMRT frequencies near the DG~Tau optical stellar position (which is denoted by the $+$ symbol in Fig.~\ref{DGTAU_323} and \ref{DGTAU_608}) even after shifting (see Section~\ref{sec:observations}). At 608\,MHz the emission is consistent with the optical stellar position indicating that it is associated with the base of the jet, however the peak of the 323\,MHz emission is approximately 8\,arcsec to the east. At 323\,MHz, the emission has $S_{\rm peak, 323\,MHz}=0.62\pm0.12$\,mJy, $S_{\rm int, 323\,MHz}=1.16\pm0.34$\,mJy and $\theta_{\rm 323\,MHz}=15.1\times4.9$\,arcsec$^2$, $61.4^\circ$. The spectral index between the GMRT measurements is therefore $\alpha_{\rm GMRT}=-1.10\pm0.72$ and when combined with the positional offset, is suggestive that the emission detected at the two frequencies may not be directly associated. Consequently, the spectral index $\alpha_{\rm GMRT}$ would be unreliable if it is extrapolated between two separate sources of emission. As a result we only report the 608\,MHz flux density and an upper limit for the flux density from the base of the outflow at 323\,MHz in Table~\ref{tab:fluxes}, and only use our measurement at 608\,MHz in the SED analysis (see Section~\ref{sec:SEDs}). 

\begin{figure}
\includegraphics[width=0.49\textwidth]{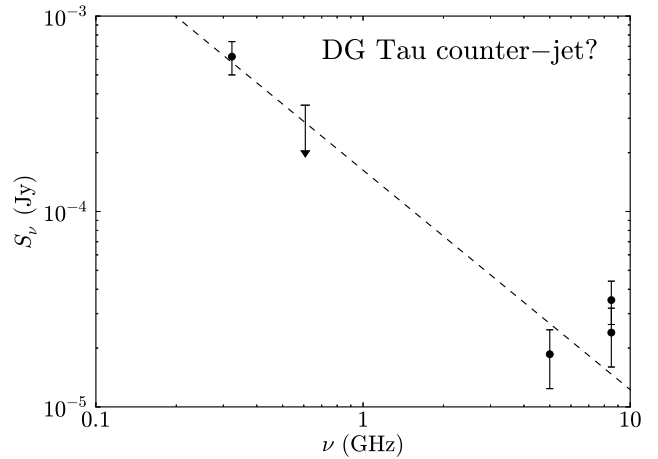}
\caption{The approximate SED for the possible DG~Tau counter-jet with $\alpha=-1.12$ denoted as a dashed line. See Section~\ref{sec:DGTAU} for more details. \label{fig:counterjet}}
\end{figure}

Unresolved emission at the $3\sigma_{\rm rms}$ level is observed with the VLA at 5.4 and 8.5\,GHz \citep{2012AA...537A.123R, 2013ApJ...766...53L} at the location of the GMRT 323\,MHz detection. Due to the unresolved nature of the emission at higher frequencies, we construct an approximate SED using the peak flux density measurements, see Fig.~\ref{fig:counterjet}. Specifically, we use $S_{\rm peak, 323\,MHz}$ from above, an upper limit at 608\,MHz of $S_{\rm peak, 608\,MHz}<0.24$\,mJy and flux densities $S_\nu=3\sigma_{{\rm rms}, \nu}\pm1\sigma_{{\rm rms}, \nu}$ from \citet{2012AA...537A.123R} and \citet{2013ApJ...766...53L} at 8.5 and 5.4\,GHz. We measure a spectral index following the method described in \citet{2014ApJ...792L..18A} and find $\alpha=-1.12\pm0.10$. This measurement should only be considered as indication of a steeply falling spectral index of order $-1$ due to the approximate nature of the SED, although it is clearly indicative of non-thermal emission. 

One possible explanation is that the emission at this location arises from a background extragalactic source, however it has been shown that the probability of detecting a background object this close in proximity to DG~Tau is very low \citep[$<10^{-3}$,][]{2014ApJ...792L..18A}. Furthermore, the significant signal to the southwest of DG~Tau also exhibits a synchrotron spectrum. It is therefore possible that the emission arising at 323\,MHz slightly offset from the optical stellar position arises from shocks in the DG~Tau redshifted counter-jet \citep[see e.g.][]{2014MNRAS.442...28W}, however more evidence is needed. 

Although we have a non-detection of free--free emission from DG~Tau at 323\,MHz, the sensitivity and resolution are too low to conclusively report the detection of the turnover frequency in the spectrum where $\tau_{\nu_0}=1$. If the turnover frequency is observed, higher resolution and sensitivity observations are required to detect the optically thick edge of the emission at these very low radio frequencies to adequately constrain the mass of the ionised gas around this YSO. We therefore use these results to place limits on these parameters in Section~\ref{sec:parameter_estimation}.

\subsection{Spectral Energy Distributions}
\label{sec:SEDs}

An extensive literature search was conducted for integrated flux densities measured at frequencies $\nu<1$\,THz ($\lambda>0.3$\,mm) and with resolution similar to the GMRT to include in the spectral energy distributions for each target source. These SEDs therefore represent these systems on scales of several arcseconds ($\sim1000$\,au). We use the previously compiled SED from \citet[][and references therein]{2012MNRAS.423.1089A} for L1551~IRS~5 and the list of archival data used in the SEDs for T~Tau and DG~Tau can be found in Appendix~\ref{appA}.

The Markov Chain Monte Carlo based Maximum Likelihood algorithm \textsc{metro} \citep{hob04} was used to fit a combined double power-law to the larger dataset of each source to model the two apparent emission components: free--free emission from the partially ionised outflow (with low frequency spectral index $\alpha$) and thermal dust emission from the circumstellar disc/envelope (with high frequency spectral index $\alpha'$) using a joint likelihood. It is important to disentangle these two emission mechanisms simultaneously as it has been shown that considering free--free and thermal dust components separately can give vastly different values for the spectral slope and normalisation of each component \citep[e.g.][]{2012MNRAS.420.3334S}. This can have implications when determining physical parameters from the free--free spectra (such as gas mass and electron density) and the thermal dust spectra (such as disc mass and grain size).

In the Rayleigh--Jeans region ($h\nu \ll k_{\rm B}T_{\rm d}$, or $\nu\ll1$\,THz for a characteristic dust temperature $T_{\rm d}=50$\,K), the thermal emission from dust grains in the circumstellar environment can be well approximated by a power-law with $S_\nu \propto \nu^{\alpha'}$ \citep[e.g.][]{2013MNRAS.435.1139S}. The spectral index $\alpha'$ of flux density measurements is related to the dust opacity index $\beta$ as $\beta \simeq (1+\Delta)\times(\alpha'-2)$, where $\Delta$ is the ratio of optically thick to optically thin emission \citep{1990AJ.....99..924B}. At long wavelengths $\Delta \rightarrow 0$ as the emission is entirely optically thin, so $\beta\approx\alpha'-2$. This allows the largest grain sizes to be determined directly from a measure of this spectral index. 

The fitted model is of the form,
\begin{equation}
\left( \frac{S_\nu}{\rm mJy} \right) = K_{323\,\rm{MHz}} \left( \frac{\nu}{\rm 323\,MHz} \right)^\alpha + K_{100\,\rm{GHz}} \left( \frac{\nu}{\rm 100\,GHz} \right)^{\alpha'},
\end{equation}
where the constants $K_{323\,\rm{MHz}}$ and $K_{100\,\rm{GHz}}$ normalise the two power-law components at 323\,MHz and 100\,GHz ($\lambda=90$\,cm and 3\,mm, respectively). Consequently, $K_{323\,\rm{MHz}}$ represents the normalised flux density at 323\,MHz (expected to be dominated by free--free emission) and $K_{100\,\rm{GHz}}$ represents the normalised flux density at 100\,GHz (expected to be dominated by thermal dust emission).  In Section~\ref{sec:parameter_estimation}, we use the derived values for $K_{323\,\rm{MHz}}$ to estimate physical parameters of the radio emitting plasma for each source.

\begin{figure}
\subfloat{\includegraphics[width=0.49\textwidth]{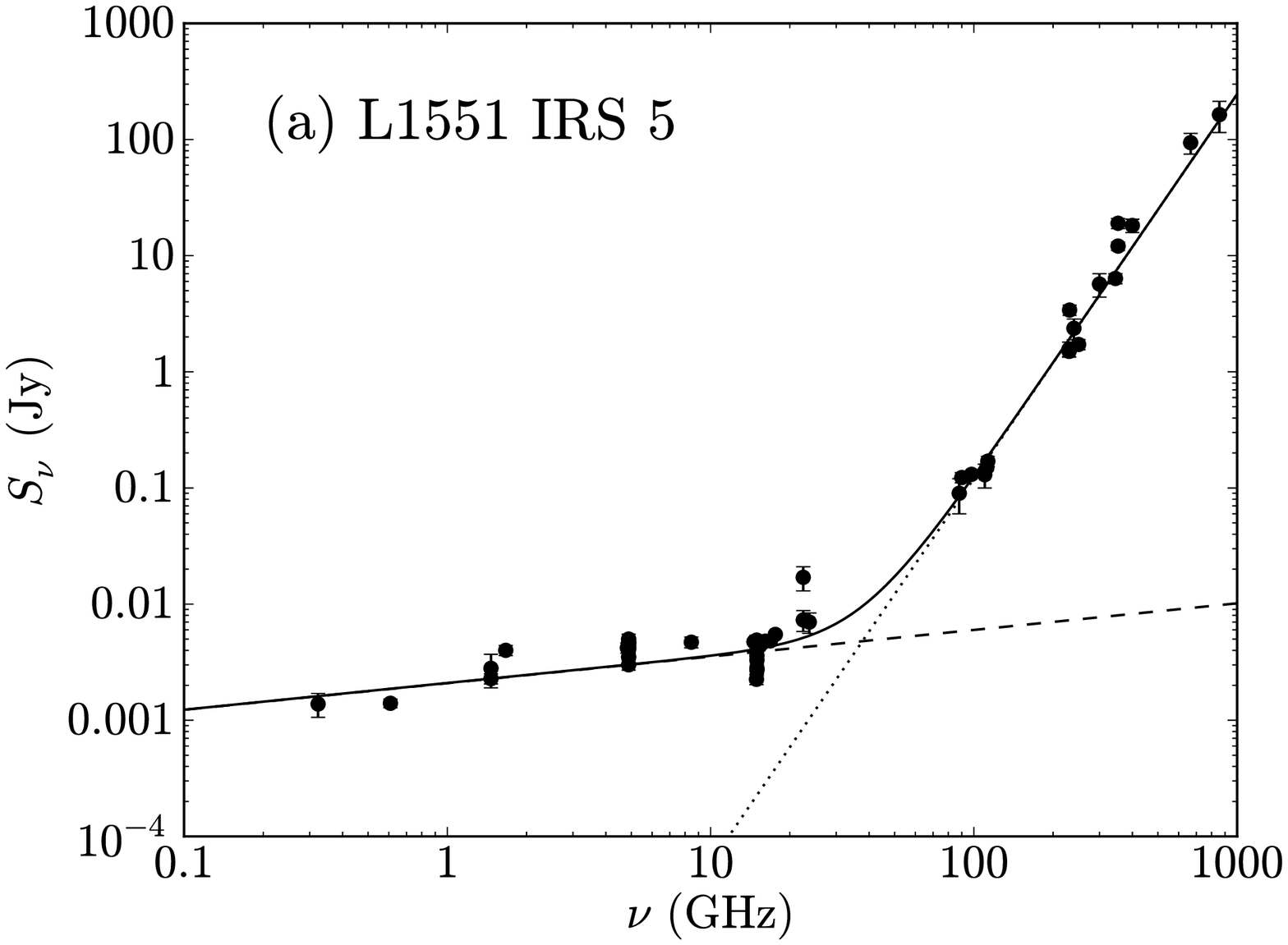} \label{L1551SED}} \\
\subfloat{\includegraphics[width=0.49\textwidth]{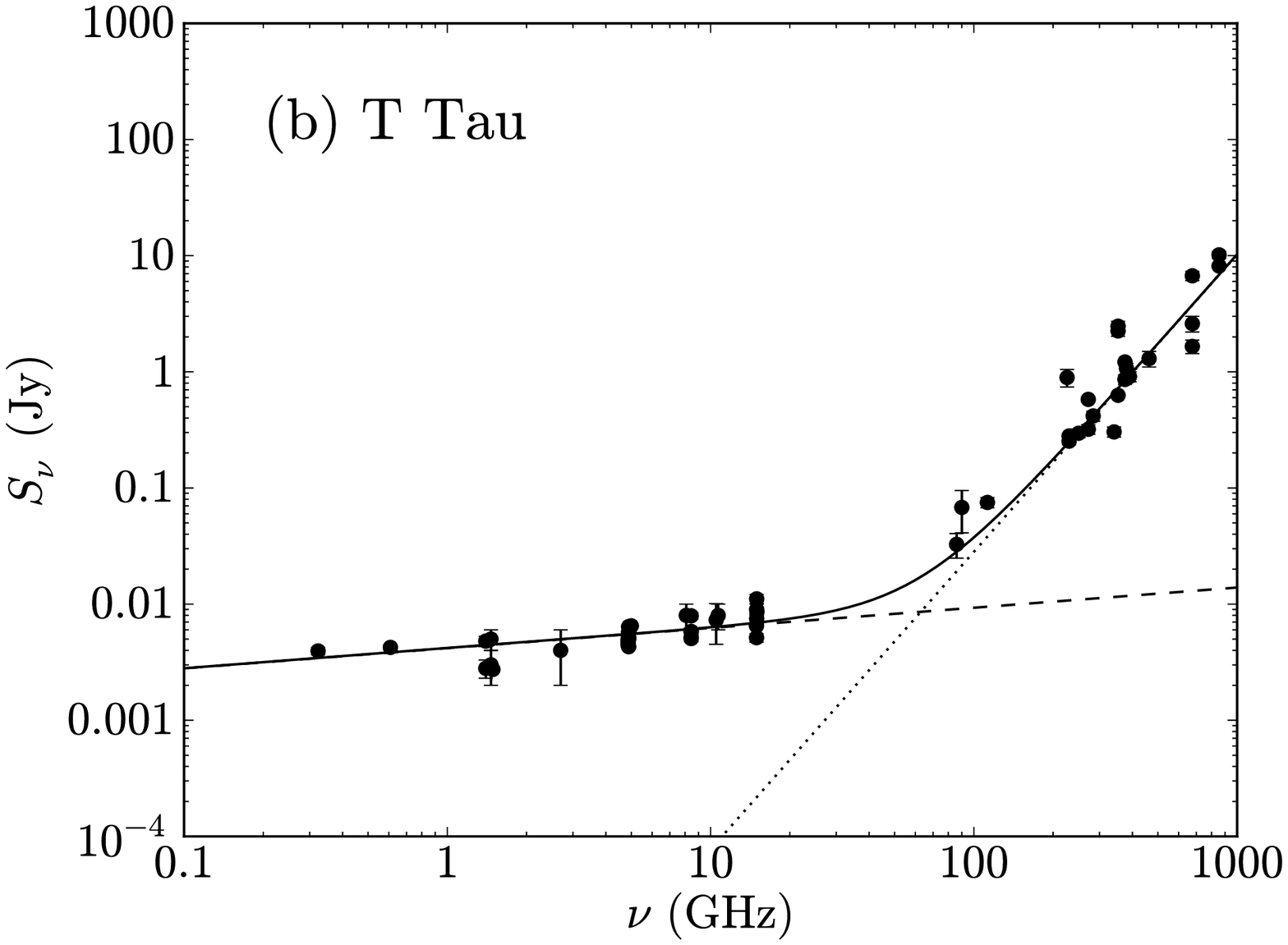} \label{TTAUSED}} \\
\subfloat{\includegraphics[width=0.49\textwidth]{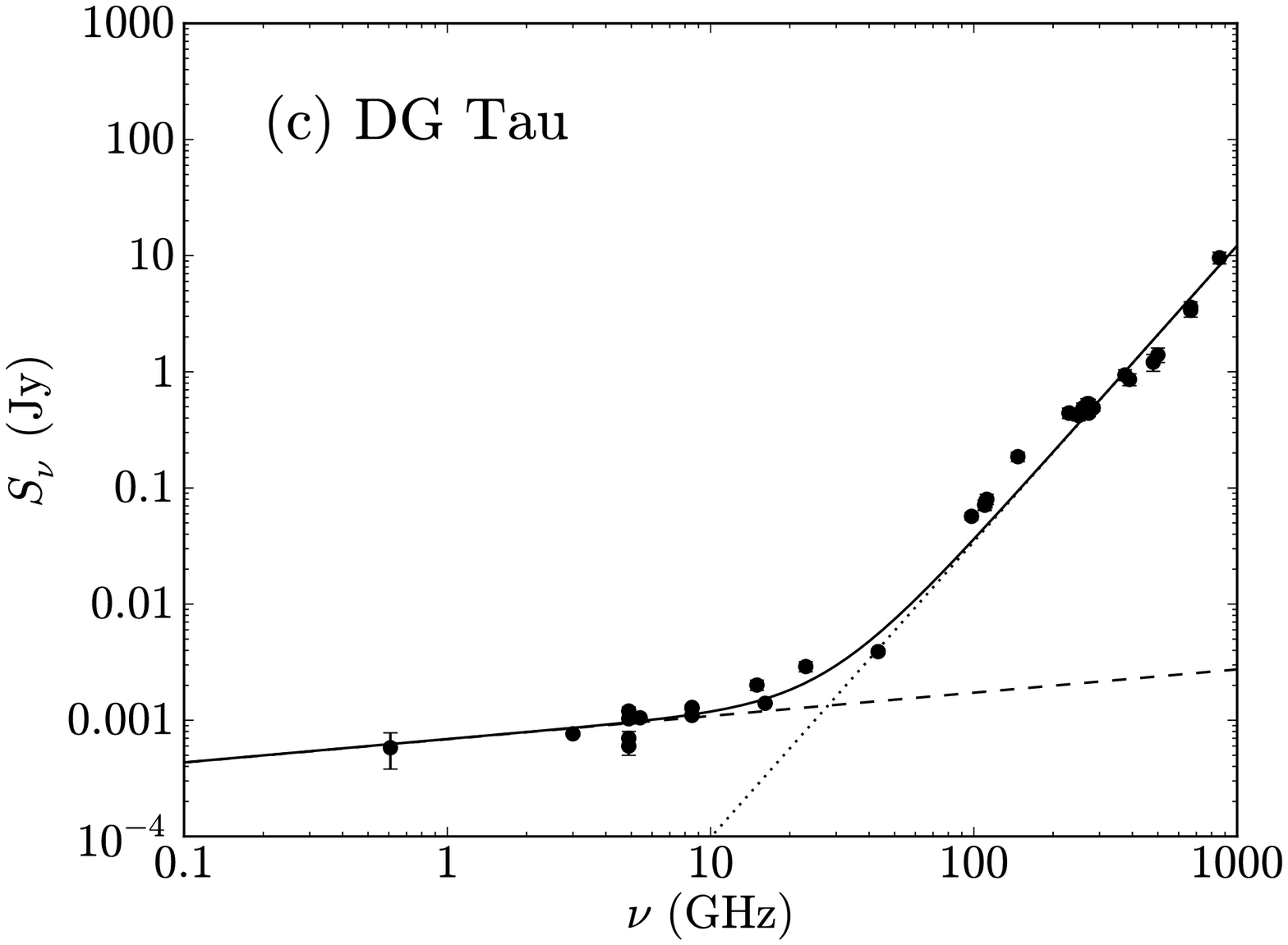} \label{DGTAUSED}}
\caption{The spectral energy distributions for (a) L1551~IRS~5, (b) T~Tau and (c) DG~Tau. Measured flux densities from the literature and this work are shown as filled circles (see Appendix~\ref{appA}). The SEDs are fitted with a combined double power-law to model the free--free emission from the outflow with spectral index $\alpha$ (dashed line) and the thermal dust emission from the circumstellar disc/envelope with high frequency spectral index $\alpha'$ (dotted line) which is related to the opacity index as $\beta\approx\alpha'-2$. The total fit is shown as a solid line. Derived parameters from the SED fitting are listed in Table~\ref{tab:SEDresults}. \label{fig:GMRTSED}}
\end{figure}

We fit all the available flux densities for each target source using uniform separable priors such that,
\begin{equation}
\Pi = \pi_{K_{323\,\rm{MHz}}}(0, 10\,{\rm mJy})~\pi_{\alpha}(-2, 2)~\pi_{K_{100\,\rm{GHz}}}(0, 1\,{\rm Jy})~\pi_{\alpha'}(0, 4). \nonumber
\end{equation}
The prior range for $\alpha$ was selected to allow a variety of possible radio emission mechanisms such as synchrotron ($\alpha\lesssim-0.7$), optically thin free--free ($\alpha\approx-0.1$) and optically thick free--free ($\alpha\approx2$). The prior range for $\alpha'$ was chosen to allow a range of values up to $\beta=2$ expected for protostellar envelopes with small, warm dust grain populations \citep[e.g.][]{2013MNRAS.435.1139S}. The prior ranges for $K_{323\,\rm{MHz}}$ and $K_{100\,\rm{GHz}}$ were chosen based on the flux densities of these objects around 323\,MHz and 100\,GHz. The SEDs with the model fits are shown in Fig.~\ref{fig:GMRTSED}, and the derived parameters from the fitting procedure are presented in Table~\ref{tab:SEDresults}.

\begin{table*}
\centering
\caption{SED modelling results. Column [1] contains the target source name; [2] the derived normalisation of the low frequency power-law at 323\,MHz; [3] the derived low frequency spectral index $\alpha$; [4] the derived normalisation of the high frequency power-law at 100\,GHz; [5] the derived opacity index $\beta$ which is related to the high frequency spectral index $\alpha'$ as $\beta\approx\alpha'-2$; [6] the geometric mean of the measured deconvolved angular source size based on the 608\,MHz data presented in Table~\ref{tab:fluxes}; [7] the estimated average electron density calculated from equation~(\ref{eq:ne}); [8] the estimated ionised gas mass calculated from equation~(\ref{eq:mi}); [9] the estimated maximum emission measure calculated from equation~(\ref{eq:em}); and [10] the estimated turnover frequency calculated from equation~(\ref{eq:v0}).}
\label{tab:SEDresults}
\begin{tabular}{lccccccccc} 
\hline
Source & $K_{323\,\rm{MHz}}$ & $\alpha$ & $K_{100\,\rm{GHz}}$ & $\beta$ & $\theta$ & $n_{\rm e}$ & $M_{\rm ion}$ & $EM$ & $\nu_0$ \\
	& (mJy) & & (mJy) & & ($''$) & (cm$^{-3}$) & (M$_{\odot}$) & (pc\,cm$^{-6}$) & (MHz)\\ 
\hline
L1551~IRS~5 	& $1.61\pm0.10$ & $0.23\pm0.02$ & $120.58\pm3.63$ & $1.31\pm0.05$ & $4.8$& $2.2\times10^{3}$ & $3.1\times10^{-6}$ & $2.3\times10^{4}$ & 105 \\
T~Tau 		& $3.43\pm0.08$ & $0.17\pm0.01$ & $28.16\pm1.15$ & $0.56\pm0.03$ & $2.9$& $6.8\times10^{3}$ & $2.2\times10^{-6}$ & $1.3\times10^{5}$ & 226 \\
DG~Tau 		& $0.55\pm0.05$ & $0.20\pm0.03$ & $34.54\pm1.08$ & $0.55\pm0.03$ & $5.3$& $1.2\times10^{3}$ & $1.8\times10^{-6}$ & $6.4\times10^{3}$ & 59 \\
\hline
\end{tabular}
\end{table*}

For all three targets, the GMRT data show a continuation of the free--free spectral index extrapolated from higher radio frequencies; the turnover frequency at which the emission becomes optically thick has not yet been reached. In addition, the low frequency spectral indices for all three sources are consistent with the value for a collimated outflow \citep[$\sim0.25$,][]{1986ApJ...304..713R} which is expected based on observations of their collimated jets at other wavelengths. In most cases, the radio emission at 323 and 608\,MHz appears resolved with the GMRT suggesting that we are observing diffuse optically thin emission  on scales of $\sim100-1000$\,au. These regions may be ionised by the UV radiation generated within the shock fronts resulting from the interaction of the outflows with the ambient medium \citep{1987RMxAA..14..595C, 1989ApL&C..27..299C, 2000ApJ...538..728G}. 

It is clear from Table~\ref{tab:SEDresults} that the Class~I and Class~II SEDs are quantitatively different. The higher value of $\beta$ derived for L1551~IRS~5 indicates that there is still a contribution to the SED from the small, warm dust grains in the circumstellar envelope. In comparison, T~Tau and DG~Tau have $\beta<1$, indicating the presence of large dust grains in their circumstellar discs and lack of a circumstellar envelope, consistent with their higher evolutionary status.

\subsection{Physical Parameter Estimation}
\label{sec:parameter_estimation}

The model of \citet{1967ApJ...147..471M} allows physical quantities such as the average electron density ($n_{\rm e}$), the mass of the ionised gas ($M_{\rm ion}$) and the maximum emission measure ($EM$) to be estimated if the flux density, apparent angular source size ($\theta$), electron temperature ($T_{\rm e}$), and distance to the source ($D$) are known. In the case of a homogeneous sphere and assuming optically thin free--free emission, the average electron density of the region can be computed using the relation
\begin{eqnarray}
\left(\frac{n_{\rm e}}{\rm cm^{-3}}\right)  =  7.2\times10^{3} \left(\frac{T_e}{10^4\,{\rm K}}\right)^{0.175} \left(\frac{\nu}{\rm GHz}\right)^{0.05} \left(\frac{S_{\nu}}{\rm mJy}\right)^{0.5} \nonumber \\
	 {}\times \left(\frac{D}{\rm kpc}\right)^{-0.5} \left(\frac{\theta}{\rm arcsec}\right)^{-1.5} \label{eq:ne}.
\end{eqnarray}
Integration of the density distribution and multiplication of the result with the ratio of the mass of a hydrogen atom to the solar mass M$_\odot$ yields the total mass of ionised hydrogen,
\begin{eqnarray}
\left(\frac{M_{\rm ion}}{\rm M_\odot}\right) =  3.4\times10^{-5} \left(\frac{T_{\rm e}}{10^4\,{\rm K}}\right)^{0.175} \left(\frac{\nu}{{\rm GHz}}\right)^{0.05} \left(\frac{S_{\nu}}{\rm mJy}\right)^{0.5} \nonumber \\
	 {}\times \left(\frac{D}{{\rm kpc}}\right)^{2.5} \left(\frac{\theta}{\rm arcsec}\right)^{1.5} \label{eq:mi}.
\end{eqnarray}
The maximum emission measure in the centre of the source is
\begin{equation}
\left(\frac{EM}{\rm pc\,cm^{-6}}\right)  =  7.1\times10^{-3} \left(\frac{D}{{\rm kpc}}\right) \left(\frac{\theta}{\rm arcsec}\right) \left(\frac{n_{\rm e}}{\rm cm^{-3}}\right)^2 \label{eq:em}.
\end{equation}

We use equations~(\ref{eq:ne})--(\ref{eq:em}) to estimate these quantities for each target YSO with the results derived from the SED fitting procedure. We adopt $S_\nu=K_{323\,\rm{MHz}}$, $\nu=323$\,MHz and angular source sizes based on the geometric mean of the deconvolved dimensions of the Gaussian fits to the higher resolution 608\,MHz data presented in Table~\ref{tab:fluxes}. We follow \citet{2013RMxAA..49...79G} and adopt an average value of $10^4$\,K for the electron temperature which is reasonable when compared with values derived from shock excited optical/infrared lines \citep[e.g.][]{1994ApJ...436..125H, 2006A&A...456..189P}. The derived $n_{\rm e}$ and $M_{\rm ion}$ do not depend strongly on the assumed $T_{\rm e}$ (given the weak dependence of $T_{\rm e}$ in equations~\ref{eq:ne} and \ref{eq:mi}); e.g. if 8000\,K is assumed instead of $10^4$\,K, then the resulting values change by a factor of only a few per~cent. As mentioned previously, the three target sources reside in the Taurus Molecular Cloud. We adopt a distances of 140\,pc to L1551~IRS~5 and T~Tau \citep{1994AJ....108.1872K} and 130\,pc to DG~Tau due to its location on the western side of the molecular cloud complex \citep{2012ApJ...747...18T}. The assumed values of $\theta$ used and the derived values for $n_{\rm e}$, $M_{\rm ion}$ and $EM$ in the case of each YSO are summarised in Table~\ref{tab:SEDresults}.

As we have not yet detected the turnover frequency in any of the free--free spectra of the target sources, the results for $M_{\rm ion}$ are upper limits and indicate that the ionised gas content around these systems is $\la3\times10^{-6}$\,M$_{\odot}$. Furthermore, there is a strong degeneracy between the parameters $M_{\rm ion}$ and $\theta$ in equation~(\ref{eq:mi}). Therefore, higher resolution observations at very low frequencies ($\nu<300$\,MHz) are required to break this degeneracy through detection of the optically thick boundary of the radio emitting plasma associated with these YSO outflows. This will allow for a more accurate measurement of the size of the emitting region which will constrain $M_{\rm ion}$ directly. It might be expected that $M_{\rm ion}$ for Class~II objects would be lower than for Class~I objects, consistent with the trend that more evolved YSOs drive less powerful outflows \citep[e.g.][]{1996AA...311..858B} and therefore would have less ionised material. We do not detect a significant difference in the values of $M_{\rm ion}$ for our small sample.

If the turnover frequency was the only limiting factor, then the estimates for $n_{\rm e}$ and $EM$ listed in Table~\ref{tab:SEDresults} would also be upper limits, however these parameters have a stronger dependence on the size of the emitting region. These values for $n_{\rm e}$ may be lower limits given the likelihood of partial ionisation, inhomogeneities and strong asymmetries as in the case of a significant contribution to the flux from jetlike geometries \citep{2000ApJ...538..728G}. If $\theta$ is smaller than the values adopted here (which is expected if the radio emission is associated with the bases of the large scale outflows), then the derived values for $n_{\rm e}$ will be lower limits. For example, using higher resolution observations at shorter wavelengths of the Class~I YSO driving HH~111, \citet{2013RMxAA..49...79G} obtain $M_{\rm ion}\lesssim2.2\times10^{-7}$\,M$_{\odot}$ and $n_{\rm e}\gtrsim4.2\times10^4$\,cm$^{-3}$ from an emitting region $\lesssim0.25$\,arcsec. 

The expected free--free turnover frequency can be estimated from the optical path length for free--free emission, which can be approximated by 
\begin{equation}
\tau_\nu = 8.235\times10^{-2} \left(\frac{T_{\rm e}}{\rm K}\right)^{-1.35} \left(\frac{\nu}{{\rm GHz}}\right)^{-2.1} \left(\frac{EM}{\rm pc\,cm^{-6}}\right)
\end{equation}
\citep{altenhoff1960}. The turnover frequency ($\nu_0$) at which the optical depth equals unity ($\tau_{\nu_0}=1$) is therefore
\begin{equation}
\left(\frac{\nu_0}{{\rm GHz}}\right) = 0.3045 \left(\frac{T_{\rm e}}{\rm K}\right)^{-0.643} \left(\frac{EM}{\rm pc\,cm^{-6}}\right)^{0.476} \label{eq:v0}.
\end{equation}
For example, an expected turnover frequency of around 665\,MHz for HH~111 is estimated based on the higher resolution results of \citet{2013RMxAA..49...79G}. We use our results for $EM$ to estimate the expected $\nu_0$ for each of the targets and list the results in Table~\ref{tab:SEDresults}. We note that these estimates are to obtain a general idea of where the turnover may take place and that more accurate measurements of source geometries are required to constrain this parameter. 

The estimation of a very low frequency turnover ($\nu_0\sim59$\,MHz) for DG~Tau suggests that we do not detect the turnover in the free--free spectrum with these GMRT data and that our non-detection of emission from the base of the outflow is likely due to a combination of low sensitivity and resolution at 323\,MHz (see Section~\ref{sec:DGTAU}). The turnover frequencies estimated for L1551~IRS~5 and T~Tau however, make these objects well suited for very long baseline observations with LOFAR ($\nu=110-250$\,MHz, sub-arcsecond resolution), particularly T~Tau which may turnover around 226\,MHz.

\section{Conclusions}
\label{sec:conclusions}

We have presented the longest wavelength investigations of low mass young stars to date. Overall, these new GMRT data show a continuation of the optically thin free--free spectrum extrapolated from shorter radio wavelengths and the turnover occurs at longer wavelengths for these objects. We use these measurements to place upper limits on the mass of the ionised gas around these systems, and find $M_{\rm ion}\la3\times10^{-6}$\,M$_{\odot}$. Higher resolution observations at longer wavelengths with instruments such as LOFAR or the VLA can be used to place constraints on the physical conditions of the ionised gas surrounding young stars, and T~Tau in particular is well-suited for further investigation. 

Our GMRT data for DG~Tau suggests that the emission detected at the two frequencies may not be directly associated. Non-thermal emission from a shock in the counter-jet may be detected at 323\,MHz, however further evidence is required. If this is indeed a non-detection of free--free emission at 323\,MHz from the base of the ionised jet, then the turnover frequency may be detected for this target and our results provide a good estimate for $M_{\rm ion}$, although our estimate for $\nu_0$ suggests otherwise. Follow-up observations with higher sensitivity and resolution can resolve this issue.

The GMRT can be used as a potential survey instrument for star forming regions at very long wavelengths due to the extent of its field of view. For example, the entirety of the L1551 star forming region fits within the 44\,arcmin FWHM of the GMRT primary beam at 608\,MHz. GMRT data can also be used to assist in the calibration of LOFAR observations towards these regions. A full survey comprising of all three fields, including a detailed description of the survey methodology and data products, will be presented in a forthcoming paper.

\section*{Acknowledgements}

We thank the staff of the GMRT who have made these observations possible. GMRT is run by the National Centre for Radio Astrophysics of the Tata Institute of Fundamental Research. REA, TPR and CPC acknowledge support from Science Foundation Ireland under grant 13/ERC/I2907. AMS gratefully acknowledges support from the European Research Council under grant  ERC-2012-StG-307215 LODESTONE. DAG thanks the Science and Technology Facilities Council for support. We thank the anonymous referee for their helpful and constructive comments to clarify this manuscript.




\bibliographystyle{mnras}
\bibliography{Bibliography} 

\begin{thebibliography}{}
\makeatletter
\relax
\def\mn@urlcharsother{\let\do\@makeother \do\$\do\&\do\#\do\^\do\_\do\%\do\~}
\def\mn@doi{\begingroup\mn@urlcharsother \@ifnextchar [ {\mn@doi@}
  {\mn@doi@[]}}
\def\mn@doi@[#1]#2{\def\@tempa{#1}\ifx\@tempa\@empty \href
  {http://dx.doi.org/#2} {doi:#2}\else \href {http://dx.doi.org/#2} {#1}\fi
  \endgroup}
\def\mn@eprint#1#2{\mn@eprint@#1:#2::\@nil}
\def\mn@eprint@arXiv#1{\href {http://arxiv.org/abs/#1} {{\tt arXiv:#1}}}
\def\mn@eprint@dblp#1{\href {http://dblp.uni-trier.de/rec/bibtex/#1.xml}
  {dblp:#1}}
\def\mn@eprint@#1:#2:#3:#4\@nil{\def\@tempa {#1}\def\@tempb {#2}\def\@tempc
  {#3}\ifx \@tempc \@empty \let \@tempc \@tempb \let \@tempb \@tempa \fi \ifx
  \@tempb \@empty \def\@tempb {arXiv}\fi \@ifundefined
  {mn@eprint@\@tempb}{\@tempb:\@tempc}{\expandafter \expandafter \csname
  mn@eprint@\@tempb\endcsname \expandafter{\@tempc}}}

\bibitem[\protect\citeauthoryear{{Adams}, {Emerson}  \& {Fuller}}{{Adams}
  et~al.}{1990}]{1990ApJ...357..606A}
{Adams} F.~C.,  {Emerson} J.~P.,   {Fuller} G.~A.,  1990, \mn@doi [\apj]
  {10.1086/168949}, \href {http://adsabs.harvard.edu/abs/1990ApJ...357..606A}
  {357, 606}

\bibitem[\protect\citeauthoryear{{Ainsworth} et~al.,}{{Ainsworth}
  et~al.}{2012}]{2012MNRAS.423.1089A}
{Ainsworth} R.~E.,  et~al., 2012, \mn@doi [\mnras]
  {10.1111/j.1365-2966.2012.20935.x}, \href
  {http://adsabs.harvard.edu/abs/2012MNRAS.423.1089A} {423, 1089}

\bibitem[\protect\citeauthoryear{{Ainsworth}, {Ray}, {Scaife}, {Greaves}  \&
  {Beswick}}{{Ainsworth} et~al.}{2013}]{2013MNRAS.436L..64A}
{Ainsworth} R.~E.,  {Ray} T.~P.,  {Scaife} A.~M.~M.,  {Greaves} J.~S.,
  {Beswick} R.~J.,  2013, \mn@doi [\mnras] {10.1093/mnrasl/slt114}, \href
  {http://adsabs.harvard.edu/abs/2013MNRAS.436L..64A} {436, L64}

\bibitem[\protect\citeauthoryear{{Ainsworth}, {Scaife}, {Ray}, {Taylor},
  {Green}  \& {Buckle}}{{Ainsworth} et~al.}{2014}]{2014ApJ...792L..18A}
{Ainsworth} R.~E.,  {Scaife} A.~M.~M.,  {Ray} T.~P.,  {Taylor} A.~M.,  {Green}
  D.~A.,   {Buckle} J.~V.,  2014, \mn@doi [\apjl]
  {10.1088/2041-8205/792/1/L18}, \href
  {http://adsabs.harvard.edu/abs/2014ApJ...792L..18A} {792, L18}

\bibitem[\protect\citeauthoryear{{Altenhoff}, {Mezger}  \&
  {Westerhout}}{{Altenhoff} et~al.}{1960}]{altenhoff1960}
{Altenhoff} W.,  {Mezger} P.~G.~{Wendker} H.,   {Westerhout} F.,  1960,
  Ver\"{o}ffentlichungen Universit\"{a}ts-Sternwarte, 59, 48

\bibitem[\protect\citeauthoryear{{Altenhoff}, {Braes}, {Olnon}  \&
  {Wendker}}{{Altenhoff} et~al.}{1976}]{1976AA....46...11A}
{Altenhoff} W.~J.,  {Braes} L.~L.~E.,  {Olnon} F.~M.,   {Wendker} H.~J.,  1976,
  \aap, \href {http://adsabs.harvard.edu/abs/1976A%26A....46...11A} {46, 11}

\bibitem[\protect\citeauthoryear{{Altenhoff}, {Huchtmeier}, {Schmidt},
  {Schraml}  \& {Stumpff}}{{Altenhoff} et~al.}{1986}]{1986AA...164..227A}
{Altenhoff} W.~J.,  {Huchtmeier} W.~K.,  {Schmidt} J.,  {Schraml} J.~B.,
  {Stumpff} P.,  1986, \aap, \href
  {http://adsabs.harvard.edu/abs/1986A%26A...164..227A} {164, 227}

\bibitem[\protect\citeauthoryear{{Altenhoff}, {Thum}  \& {Wendker}}{{Altenhoff}
  et~al.}{1994}]{1994AA...281..161A}
{Altenhoff} W.~J.,  {Thum} C.,   {Wendker} H.~J.,  1994, \aap, \href
  {http://cdsads.u-strasbg.fr/abs/1994A%26A...281..161A} {281, 161}

\bibitem[\protect\citeauthoryear{{Ananthakrishnan}}{{Ananthakrishnan}}{2005}]{2005ICRC...10..125A}
{Ananthakrishnan} S.,  2005, International Cosmic Ray Conference, \href
  {http://adsabs.harvard.edu/abs/2005ICRC...10..125A} {10, 125}

\bibitem[\protect\citeauthoryear{{Andr{\'e}}, {Ward-Thompson}  \&
  {Barsony}}{{Andr{\'e}} et~al.}{2000}]{2000prpl.conf...59A}
{Andr{\'e}} P.,  {Ward-Thompson} D.,   {Barsony} M.,  2000, Protostars and
  Planets IV, \href {http://adsabs.harvard.edu/abs/2000prpl.conf...59A} {p.~59}

\bibitem[\protect\citeauthoryear{{Andrews} \& {Williams}}{{Andrews} \&
  {Williams}}{2005}]{2005ApJ...631.1134A}
{Andrews} S.~M.,  {Williams} J.~P.,  2005, \mn@doi [\apj] {10.1086/432712},
  \href {http://adsabs.harvard.edu/abs/2005ApJ...631.1134A} {631, 1134}

\bibitem[\protect\citeauthoryear{{Anglada}, {Rodr{\'{\i}}guez}  \&
  {Carrasco-Gonzalez}}{{Anglada} et~al.}{2015}]{2015aska.confE.121A}
{Anglada} G.,  {Rodr{\'{\i}}guez} L.~F.,   {Carrasco-Gonzalez} C.,  2015,
  Advancing Astrophysics with the Square Kilometre Array (AASKA14), \href
  {http://adsabs.harvard.edu/abs/2015aska.confE.121A} {p.~121}

\bibitem[\protect\citeauthoryear{{Beckwith} \& {Sargent}}{{Beckwith} \&
  {Sargent}}{1991}]{1991ApJ...381..250B}
{Beckwith} S.~V.~W.,  {Sargent} A.~I.,  1991, \mn@doi [\apj] {10.1086/170646},
  \href {http://adsabs.harvard.edu/abs/1991ApJ...381..250B} {381, 250}

\bibitem[\protect\citeauthoryear{{Beckwith}, {Sargent}, {Chini}  \&
  {Guesten}}{{Beckwith} et~al.}{1990}]{1990AJ.....99..924B}
{Beckwith} S.~V.~W.,  {Sargent} A.~I.,  {Chini} R.~S.,   {Guesten} R.,  1990,
  \mn@doi [\aj] {10.1086/115385}, \href
  {http://adsabs.harvard.edu/abs/1990AJ.....99..924B} {99, 924}

\bibitem[\protect\citeauthoryear{{Bertout} \& {Thum}}{{Bertout} \&
  {Thum}}{1982}]{1982AA...107..368B}
{Bertout} C.,  {Thum} C.,  1982, \aap, \href
  {http://adsabs.harvard.edu/abs/1982A%26A...107..368B} {107, 368}

\bibitem[\protect\citeauthoryear{{Bieging} \& {Cohen}}{{Bieging} \&
  {Cohen}}{1985}]{1985ApJ...289L...5B}
{Bieging} J.~H.,  {Cohen} M.,  1985, \mn@doi [\apjl] {10.1086/184423}, \href
  {http://cdsads.u-strasbg.fr/abs/1985ApJ...289L...5B} {289, L5}

\bibitem[\protect\citeauthoryear{{Bieging}, {Cohen}  \& {Schwartz}}{{Bieging}
  et~al.}{1984}]{1984ApJ...282..699B}
{Bieging} J.~H.,  {Cohen} M.,   {Schwartz} P.~R.,  1984, \mn@doi [\apj]
  {10.1086/162251}, \href {http://cdsads.u-strasbg.fr/abs/1984ApJ...282..699B}
  {282, 699}

\bibitem[\protect\citeauthoryear{{Bontemps}, {Andr{\'e}}, {Terebey}  \&
  {Cabrit}}{{Bontemps} et~al.}{1996}]{1996AA...311..858B}
{Bontemps} S.,  {Andr{\'e}} P.,  {Terebey} S.,   {Cabrit} S.,  1996, \aap,
  \href {http://adsabs.harvard.edu/abs/1996A%26A...311..858B} {311, 858}

\bibitem[\protect\citeauthoryear{{Carrasco-Gonz{\'a}lez}, {Rodr{\'{\i}}guez},
  {Anglada}, {Mart{\'{\i}}}, {Torrelles}  \& {Osorio}}{{Carrasco-Gonz{\'a}lez}
  et~al.}{2010}]{2010Sci...330.1209C}
{Carrasco-Gonz{\'a}lez} C.,  {Rodr{\'{\i}}guez} L.~F.,  {Anglada} G.,
  {Mart{\'{\i}}} J.,  {Torrelles} J.~M.,   {Osorio} M.,  2010, \mn@doi
  [Science] {10.1126/science.1195589}, \href
  {http://adsabs.harvard.edu/abs/2010Sci...330.1209C} {330, 1209}

\bibitem[\protect\citeauthoryear{{Cohen} \& {Bieging}}{{Cohen} \&
  {Bieging}}{1986}]{1986AJ.....92.1396C}
{Cohen} M.,  {Bieging} J.~H.,  1986, \mn@doi [\aj] {10.1086/114273}, \href
  {http://adsabs.harvard.edu/abs/1986AJ.....92.1396C} {92, 1396}

\bibitem[\protect\citeauthoryear{{Cohen}, {Bieging}  \& {Schwartz}}{{Cohen}
  et~al.}{1982}]{1982ApJ...253..707C}
{Cohen} M.,  {Bieging} J.~H.,   {Schwartz} P.~R.,  1982, \mn@doi [\apj]
  {10.1086/159671}, \href {http://cdsads.u-strasbg.fr/abs/1982ApJ...253..707C}
  {253, 707}

\bibitem[\protect\citeauthoryear{{Condon}, {Cotton}, {Greisen}, {Yin},
  {Perley}, {Taylor}  \& {Broderick}}{{Condon}
  et~al.}{1998}]{1998AJ....115.1693C}
{Condon} J.~J.,  {Cotton} W.~D.,  {Greisen} E.~W.,  {Yin} Q.~F.,  {Perley}
  R.~A.,  {Taylor} G.~B.,   {Broderick} J.~J.,  1998, \mn@doi [\aj]
  {10.1086/300337}, \href {http://adsabs.harvard.edu/abs/1998AJ....115.1693C}
  {115, 1693}

\bibitem[\protect\citeauthoryear{{Curiel}}{{Curiel}}{1995}]{1995RMxAC...1...59C}
{Curiel} S.,  1995, in Revista Mexicana de Astronom\'{\i}a y Astrof\'{\i}sica
  Conference Series.

\bibitem[\protect\citeauthoryear{{Curiel}, {Canto}  \&
  {Rodr{\'{\i}}guez}}{{Curiel} et~al.}{1987}]{1987RMxAA..14..595C}
{Curiel} S.,  {Canto} J.,   {Rodr{\'{\i}}guez} L.~F.,  1987, \rmxaa, \href
  {http://adsabs.harvard.edu/abs/1987RMxAA..14..595C} {14, 595}

\bibitem[\protect\citeauthoryear{{Curiel}, {Rodr{\'{\i}}guez}, {Bohigas},
  {Roth}, {Canto}  \& {Torrelles}}{{Curiel} et~al.}{1989}]{1989ApL&C..27..299C}
{Curiel} S.,  {Rodr{\'{\i}}guez} L.~F.,  {Bohigas} J.,  {Roth} M.,  {Canto} J.,
    {Torrelles} J.~M.,  1989, Astrophysical Letters and Communications, \href
  {http://adsabs.harvard.edu/abs/1989ApL%26C..27..299C} {27, 299}

\bibitem[\protect\citeauthoryear{{Curiel}, {Rodr{\'{\i}}guez}, {Moran}  \&
  {Canto}}{{Curiel} et~al.}{1993}]{1993ApJ...415..191C}
{Curiel} S.,  {Rodr{\'{\i}}guez} L.~F.,  {Moran} J.~M.,   {Canto} J.,  1993,
  \mn@doi [\apj] {10.1086/173155}, \href
  {http://adsabs.harvard.edu/abs/1993ApJ...415..191C} {415, 191}

\bibitem[\protect\citeauthoryear{{Di~Francesco}, {Johnstone}, {Kirk},
  {MacKenzie}  \& {Ledwosinska}}{{Di~Francesco}
  et~al.}{2008}]{2008ApJS..175..277D}
{Di~Francesco} J.,  {Johnstone} D.,  {Kirk} H.,  {MacKenzie} T.,
  {Ledwosinska} E.,  2008, \mn@doi [\apjs] {10.1086/523645}, \href
  {http://adsabs.harvard.edu/abs/2008ApJS..175..277D} {175, 277}

\bibitem[\protect\citeauthoryear{{Dougados}, {Cabrit}, {Lavalley}  \&
  {M{\'e}nard}}{{Dougados} et~al.}{2000}]{2000A&A...357L..61D}
{Dougados} C.,  {Cabrit} S.,  {Lavalley} C.,   {M{\'e}nard} F.,  2000, \aap,
  \href {http://adsabs.harvard.edu/abs/2000A%26A...357L..61D} {357, L61}

\bibitem[\protect\citeauthoryear{{Dyck}, {Simon}  \& {Zuckerman}}{{Dyck}
  et~al.}{1982}]{1982ApJ...255L.103D}
{Dyck} H.~M.,  {Simon} T.,   {Zuckerman} B.,  1982, \mn@doi [\apjl]
  {10.1086/183778}, \href {http://adsabs.harvard.edu/abs/1982ApJ...255L.103D}
  {255, L103}

\bibitem[\protect\citeauthoryear{{Dzib} et~al.,}{{Dzib}
  et~al.}{2013}]{2013ApJ...775...63D}
{Dzib} S.~A.,  et~al., 2013, \mn@doi [\apj] {10.1088/0004-637X/775/1/63}, \href
  {http://adsabs.harvard.edu/abs/2013ApJ...775...63D} {775, 63}

\bibitem[\protect\citeauthoryear{{Dzib} et~al.,}{{Dzib}
  et~al.}{2015}]{2015ApJ...801...91D}
{Dzib} S.~A.,  et~al., 2015, \mn@doi [\apj] {10.1088/0004-637X/801/2/91}, \href
  {http://adsabs.harvard.edu/abs/2015ApJ...801...91D} {801, 91}

\bibitem[\protect\citeauthoryear{{Eisl{\"o}ffel} \& {Mundt}}{{Eisl{\"o}ffel} \&
  {Mundt}}{1998}]{1998AJ....115.1554E}
{Eisl{\"o}ffel} J.,  {Mundt} R.,  1998, \mn@doi [\aj] {10.1086/300282}, \href
  {http://adsabs.harvard.edu/abs/1998AJ....115.1554E} {115, 1554}

\bibitem[\protect\citeauthoryear{{Evans}, {Levreault}, {Beckwith}  \&
  {Skrutskie}}{{Evans} et~al.}{1987}]{1987ApJ...320..364E}
{Evans} II N.~J.,  {Levreault} R.~M.,  {Beckwith} S.,   {Skrutskie} M.,  1987,
  \mn@doi [\apj] {10.1086/165550}, \href
  {http://cdsads.u-strasbg.fr/abs/1987ApJ...320..364E} {320, 364}

\bibitem[\protect\citeauthoryear{{Frank} et~al.,}{{Frank}
  et~al.}{2014}]{2014arXiv1402.3553F}
{Frank} A.,  et~al., 2014, preprint, \href
  {http://adsabs.harvard.edu/abs/2014arXiv1402.3553F} {} (\mn@eprint {arXiv}
  {1402.3553})

\bibitem[\protect\citeauthoryear{{Fridlund}, {Bergman}, {White}, {Pilbratt}  \&
  {Tauber}}{{Fridlund} et~al.}{2002}]{2002A&A...382..573F}
{Fridlund} C.~V.~M.,  {Bergman} P.,  {White} G.~J.,  {Pilbratt} G.~L.,
  {Tauber} J.~A.,  2002, \mn@doi [\aap] {10.1051/0004-6361:20011519}, \href
  {http://adsabs.harvard.edu/abs/2002A%26A...382..573F} {382, 573}

\bibitem[\protect\citeauthoryear{{Garn}, {Green}, {Hales}, {Riley}  \&
  {Alexander}}{{Garn} et~al.}{2007}]{2007MNRAS.376.1251G}
{Garn} T.,  {Green} D.~A.,  {Hales} S.~E.~G.,  {Riley} J.~M.,   {Alexander} P.,
   2007, \mn@doi [\mnras] {10.1111/j.1365-2966.2007.11514.x}, \href
  {http://adsabs.harvard.edu/abs/2007MNRAS.376.1251G} {376, 1251}

\bibitem[\protect\citeauthoryear{{Ghez}, {Neugebauer}, {Gorham}, {Haniff},
  {Kulkarni}, {Matthews}, {Koresko}  \& {Beckwith}}{{Ghez}
  et~al.}{1991}]{1991AJ....102.2066G}
{Ghez} A.~M.,  {Neugebauer} G.,  {Gorham} P.~W.,  {Haniff} C.~A.,  {Kulkarni}
  S.~R.,  {Matthews} K.,  {Koresko} C.,   {Beckwith} S.,  1991, \mn@doi [\aj]
  {10.1086/116030}, \href {http://adsabs.harvard.edu/abs/1991AJ....102.2066G}
  {102, 2066}

\bibitem[\protect\citeauthoryear{{Giovanardi}, {Rodr{\'{\i}}guez}, {Lizano}  \&
  {Cant{\'o}}}{{Giovanardi} et~al.}{2000}]{2000ApJ...538..728G}
{Giovanardi} C.,  {Rodr{\'{\i}}guez} L.~F.,  {Lizano} S.,   {Cant{\'o}} J.,
  2000, \mn@doi [\apj] {10.1086/309138}, \href
  {http://adsabs.harvard.edu/abs/2000ApJ...538..728G} {538, 728}

\bibitem[\protect\citeauthoryear{{Girart}, {Curiel}, {Rodr{\'{\i}}guez}  \&
  {Cant{\'o}}}{{Girart} et~al.}{2002}]{2002RMxAA..38..169G}
{Girart} J.~M.,  {Curiel} S.,  {Rodr{\'{\i}}guez} L.~F.,   {Cant{\'o}} J.,
  2002, \rmxaa, \href {http://adsabs.harvard.edu/abs/2002RMxAA..38..169G} {38,
  169}

\bibitem[\protect\citeauthoryear{{G{\'o}mez}, {Rodr{\'{\i}}guez}  \&
  {Loinard}}{{G{\'o}mez} et~al.}{2013}]{2013RMxAA..49...79G}
{G{\'o}mez} L.,  {Rodr{\'{\i}}guez} L.~F.,   {Loinard} L.,  2013, \rmxaa, \href
  {http://adsabs.harvard.edu/abs/2013RMxAA..49...79G} {49, 79}

\bibitem[\protect\citeauthoryear{{G{\"u}del}, {Skinner}, {Audard}, {Briggs}  \&
  {Cabrit}}{{G{\"u}del} et~al.}{2008}]{2008A&A...478..797G}
{G{\"u}del} M.,  {Skinner} S.~L.,  {Audard} M.,  {Briggs} K.~R.,   {Cabrit} S.,
   2008, \mn@doi [\aap] {10.1051/0004-6361:20078141}, \href
  {http://adsabs.harvard.edu/abs/2008A%26A...478..797G} {478, 797}

\bibitem[\protect\citeauthoryear{{Gustafsson}, {Kristensen}, {Kasper}  \&
  {Herbst}}{{Gustafsson} et~al.}{2010}]{2010A&A...517A..19G}
{Gustafsson} M.,  {Kristensen} L.~E.,  {Kasper} M.,   {Herbst} T.~M.,  2010,
  \mn@doi [\aap] {10.1051/0004-6361/200913828}, \href
  {http://adsabs.harvard.edu/abs/2010A%26A...517A..19G} {517, A19}

\bibitem[\protect\citeauthoryear{{Harris}, {Andrews}, {Wilner}  \&
  {Kraus}}{{Harris} et~al.}{2012}]{2012ApJ...751..115H}
{Harris} R.~J.,  {Andrews} S.~M.,  {Wilner} D.~J.,   {Kraus} A.~L.,  2012,
  \mn@doi [\apj] {10.1088/0004-637X/751/2/115}, \href
  {http://adsabs.harvard.edu/abs/2012ApJ...751..115H} {751, 115}

\bibitem[\protect\citeauthoryear{{Hartigan}, {Morse}  \& {Raymond}}{{Hartigan}
  et~al.}{1994}]{1994ApJ...436..125H}
{Hartigan} P.,  {Morse} J.~A.,   {Raymond} J.,  1994, \mn@doi [\apj]
  {10.1086/174887}, \href {http://adsabs.harvard.edu/abs/1994ApJ...436..125H}
  {436, 125}

\bibitem[\protect\citeauthoryear{{Herbst}, {Robberto}  \& {Beckwith}}{{Herbst}
  et~al.}{1997}]{1997AJ....114..744H}
{Herbst} T.~M.,  {Robberto} M.,   {Beckwith} S.~V.~W.,  1997, \mn@doi [\aj]
  {10.1086/118508}, \href {http://adsabs.harvard.edu/abs/1997AJ....114..744H}
  {114, 744}

\bibitem[\protect\citeauthoryear{{Hobson} \& {Baldwin}}{{Hobson} \&
  {Baldwin}}{2004}]{hob04}
{Hobson} M.~P.,  {Baldwin} J.~E.,  2004, \ao, 43, 2651

\bibitem[\protect\citeauthoryear{{Jewitt}}{{Jewitt}}{1994}]{1994AJ....108..661J}
{Jewitt} D.~C.,  1994, \mn@doi [\aj] {10.1086/117101}, \href
  {http://adsabs.harvard.edu/abs/1994AJ....108..661J} {108, 661}

\bibitem[\protect\citeauthoryear{{Johnston}, {Gaume}, {Fey}, {de Vegt}  \&
  {Claussen}}{{Johnston} et~al.}{2003}]{2003AJ....125..858J}
{Johnston} K.~J.,  {Gaume} R.~A.,  {Fey} A.~L.,  {de Vegt} C.,   {Claussen}
  M.~J.,  2003, \mn@doi [\aj] {10.1086/345727}, \href
  {http://adsabs.harvard.edu/abs/2003AJ....125..858J} {125, 858}

\bibitem[\protect\citeauthoryear{{Joy}}{{Joy}}{1945}]{1945ApJ...102..168J}
{Joy} A.~H.,  1945, \mn@doi [\apj] {10.1086/144749}, \href
  {http://adsabs.harvard.edu/abs/1945ApJ...102..168J} {102, 168}

\bibitem[\protect\citeauthoryear{{Kenyon}, {Dobrzycka}  \& {Hartmann}}{{Kenyon}
  et~al.}{1994}]{1994AJ....108.1872K}
{Kenyon} S.~J.,  {Dobrzycka} D.,   {Hartmann} L.,  1994, \mn@doi [\aj]
  {10.1086/117200}, \href {http://adsabs.harvard.edu/abs/1994AJ....108.1872K}
  {108, 1872}

\bibitem[\protect\citeauthoryear{{Kitamura}, {Kawabe}  \& {Saito}}{{Kitamura}
  et~al.}{1996a}]{1996ApJ...457..277K}
{Kitamura} Y.,  {Kawabe} R.,   {Saito} M.,  1996a, \mn@doi [\apj]
  {10.1086/176728}, \href {http://adsabs.harvard.edu/abs/1996ApJ...457..277K}
  {457, 277}

\bibitem[\protect\citeauthoryear{{Kitamura}, {Kawabe}  \& {Saito}}{{Kitamura}
  et~al.}{1996b}]{1996ApJ...465L.137K}
{Kitamura} Y.,  {Kawabe} R.,   {Saito} M.,  1996b, \mn@doi [\apjl]
  {10.1086/310152}, \href {http://adsabs.harvard.edu/abs/1996ApJ...465L.137K}
  {465, L137}

\bibitem[\protect\citeauthoryear{{Koresko}}{{Koresko}}{2000}]{2000ApJ...531L.147K}
{Koresko} C.~D.,  2000, \mn@doi [\apjl] {10.1086/312543}, \href
  {http://adsabs.harvard.edu/abs/2000ApJ...531L.147K} {531, L147}

\bibitem[\protect\citeauthoryear{{Ku}}{{Ku}}{1966}]{Ku1966}
{Ku} H.~H.,  1966, Journal of Research of the National Bureau of Standards,
  70C, 263

\bibitem[\protect\citeauthoryear{{Lavalley}, {Cabrit}, {Dougados}, {Ferruit}
  \& {Bacon}}{{Lavalley} et~al.}{1997}]{1997A&A...327..671L}
{Lavalley} C.,  {Cabrit} S.,  {Dougados} C.,  {Ferruit} P.,   {Bacon} R.,
  1997, \aap, \href {http://adsabs.harvard.edu/abs/1997A%26A...327..671L} {327,
  671}

\bibitem[\protect\citeauthoryear{{Lim} \& {Takakuwa}}{{Lim} \&
  {Takakuwa}}{2006}]{2006ApJ...653..425L}
{Lim} J.,  {Takakuwa} S.,  2006, \mn@doi [\apj] {10.1086/508510}, \href
  {http://adsabs.harvard.edu/abs/2006ApJ...653..425L} {653, 425}

\bibitem[\protect\citeauthoryear{{Loinard}, {Rodr{\'{\i}}guez}, {D'Alessio},
  {Rodr{\'{\i}}guez}  \& {Gonz{\'a}lez}}{{Loinard}
  et~al.}{2007}]{2007ApJ...657..916L}
{Loinard} L.,  {Rodr{\'{\i}}guez} L.~F.,  {D'Alessio} P.,  {Rodr{\'{\i}}guez}
  M.~I.,   {Gonz{\'a}lez} R.~F.,  2007, \mn@doi [\apj] {10.1086/510994}, \href
  {http://adsabs.harvard.edu/abs/2007ApJ...657..916L} {657, 916}

\bibitem[\protect\citeauthoryear{{Lynch}, {Mutel}, {G{\"u}del}, {Ray},
  {Skinner}, {Schneider}  \& {Gayley}}{{Lynch}
  et~al.}{2013}]{2013ApJ...766...53L}
{Lynch} C.,  {Mutel} R.~L.,  {G{\"u}del} M.,  {Ray} T.,  {Skinner} S.~L.,
  {Schneider} P.~C.,   {Gayley} K.~G.,  2013, \mn@doi [\apj]
  {10.1088/0004-637X/766/1/53}, \href
  {http://adsabs.harvard.edu/abs/2013ApJ...766...53L} {766, 53}

\bibitem[\protect\citeauthoryear{{Maran}, {Brown}, {Hobbs}, {Jura}  \&
  {Knapp}}{{Maran} et~al.}{1979}]{1979AJ.....84.1709M}
{Maran} S.~P.,  {Brown} R.~L.,  {Hobbs} R.~W.,  {Jura} M.,   {Knapp} G.~R.,
  1979, \mn@doi [\aj] {10.1086/112599}, \href
  {http://adsabs.harvard.edu/abs/1979AJ.....84.1709M} {84, 1709}

\bibitem[\protect\citeauthoryear{{McGroarty} \& {Ray}}{{McGroarty} \&
  {Ray}}{2004}]{2004A&A...420..975M}
{McGroarty} F.,  {Ray} T.~P.,  2004, \mn@doi [\aap]
  {10.1051/0004-6361:20041124}, \href
  {http://adsabs.harvard.edu/abs/2004A%26A...420..975M} {420, 975}

\bibitem[\protect\citeauthoryear{{Mezger} \& {Henderson}}{{Mezger} \&
  {Henderson}}{1967}]{1967ApJ...147..471M}
{Mezger} P.~G.,  {Henderson} A.~P.,  1967, \mn@doi [\apj] {10.1086/149030},
  \href {http://adsabs.harvard.edu/abs/1967ApJ...147..471M} {147, 471}

\bibitem[\protect\citeauthoryear{{Mundt} \& {Fried}}{{Mundt} \&
  {Fried}}{1983}]{1983ApJ...274L..83M}
{Mundt} R.,  {Fried} J.~W.,  1983, \mn@doi [\apjl] {10.1086/184155}, \href
  {http://adsabs.harvard.edu/abs/1983ApJ...274L..83M} {274, L83}

\bibitem[\protect\citeauthoryear{{Ohashi}, {Kawabe}, {Ishiguro}  \&
  {Hayashi}}{{Ohashi} et~al.}{1991}]{1991AJ....102.2054O}
{Ohashi} N.,  {Kawabe} R.,  {Ishiguro} M.,   {Hayashi} M.,  1991, \mn@doi [\aj]
  {10.1086/116029}, \href {http://cdsads.u-strasbg.fr/abs/1991AJ....102.2054O}
  {102, 2054}

\bibitem[\protect\citeauthoryear{{Panagia} \& {Felli}}{{Panagia} \&
  {Felli}}{1975}]{1975A&A....39....1P}
{Panagia} N.,  {Felli} M.,  1975, \aap, \href
  {http://adsabs.harvard.edu/abs/1975A%26A....39....1P} {39, 1}

\bibitem[\protect\citeauthoryear{{Perley} \& {Butler}}{{Perley} \&
  {Butler}}{2013}]{2013ApJS..204...19P}
{Perley} R.~A.,  {Butler} B.~J.,  2013, \mn@doi [\apjs]
  {10.1088/0067-0049/204/2/19}, \href
  {http://adsabs.harvard.edu/abs/2013ApJS..204...19P} {204, 19}

\bibitem[\protect\citeauthoryear{{Podio}, {Bacciotti}, {Nisini},
  {Eisl{\"o}ffel}, {Massi}, {Giannini}  \& {Ray}}{{Podio}
  et~al.}{2006}]{2006A&A...456..189P}
{Podio} L.,  {Bacciotti} F.,  {Nisini} B.,  {Eisl{\"o}ffel} J.,  {Massi} F.,
  {Giannini} T.,   {Ray} T.~P.,  2006, \mn@doi [\aap]
  {10.1051/0004-6361:20054156}, \href
  {http://adsabs.harvard.edu/abs/2006A%26A...456..189P} {456, 189}

\bibitem[\protect\citeauthoryear{{Pravdo}, {Feigelson}, {Garmire}, {Maeda},
  {Tsuboi}  \& {Bally}}{{Pravdo} et~al.}{2001}]{2001Natur.413..708P}
{Pravdo} S.~H.,  {Feigelson} E.~D.,  {Garmire} G.,  {Maeda} Y.,  {Tsuboi} Y.,
  {Bally} J.,  2001, \mn@doi [\nat] {10.1038/35099508}, \href
  {http://adsabs.harvard.edu/abs/2001Natur.413..708P} {413, 708}

\bibitem[\protect\citeauthoryear{{Pyo}, {Hayashi}, {Kobayashi}, {Terada}  \&
  {Tokunaga}}{{Pyo} et~al.}{2009}]{2009ApJ...694..654P}
{Pyo} T.-S.,  {Hayashi} M.,  {Kobayashi} N.,  {Terada} H.,   {Tokunaga} A.~T.,
  2009, \mn@doi [\apj] {10.1088/0004-637X/694/1/654}, \href
  {http://adsabs.harvard.edu/abs/2009ApJ...694..654P} {694, 654}

\bibitem[\protect\citeauthoryear{{Ray}, {Muxlow}, {Axon}, {Brown}, {Corcoran},
  {Dyson}  \& {Mundt}}{{Ray} et~al.}{1997}]{1997Natur.385..415R}
{Ray} T.~P.,  {Muxlow} T.~W.~B.,  {Axon} D.~J.,  {Brown} A.,  {Corcoran} D.,
  {Dyson} J.,   {Mundt} R.,  1997, \mn@doi [\nat] {10.1038/385415a0}, \href
  {http://adsabs.harvard.edu/abs/1997Natur.385..415R} {385, 415}

\bibitem[\protect\citeauthoryear{{Reipurth}, {Chini}, {Krugel}, {Kreysa}  \&
  {Sievers}}{{Reipurth} et~al.}{1993}]{1993AA...273..221R}
{Reipurth} B.,  {Chini} R.,  {Krugel} E.,  {Kreysa} E.,   {Sievers} A.,  1993,
  \aap, \href {http://adsabs.harvard.edu/abs/1993A%26A...273..221R} {273, 221}

\bibitem[\protect\citeauthoryear{{Reipurth}, {Bally}  \& {Devine}}{{Reipurth}
  et~al.}{1997}]{1997AJ....114.2708R}
{Reipurth} B.,  {Bally} J.,   {Devine} D.,  1997, \mn@doi [\aj]
  {10.1086/118681}, \href {http://adsabs.harvard.edu/abs/1997AJ....114.2708R}
  {114, 2708}

\bibitem[\protect\citeauthoryear{{Reynolds}}{{Reynolds}}{1986}]{1986ApJ...304..713R}
{Reynolds} S.~P.,  1986, \mn@doi [\apj] {10.1086/164209}, \href
  {http://adsabs.harvard.edu/abs/1986ApJ...304..713R} {304, 713}

\bibitem[\protect\citeauthoryear{{Rodr{\'{\i}}guez} \&
  {Canto}}{{Rodr{\'{\i}}guez} \& {Canto}}{1983}]{1983RMxAA...8..163R}
{Rodr{\'{\i}}guez} L.~F.,  {Canto} J.,  1983, \rmxaa, \href
  {http://adsabs.harvard.edu/abs/1983RMxAA...8..163R} {8, 163}

\bibitem[\protect\citeauthoryear{{Rodr{\'{\i}}guez} et~al.,}{{Rodr{\'{\i}}guez}
  et~al.}{1998}]{1998Natur.395..355R}
{Rodr{\'{\i}}guez} L.~F.,  et~al., 1998, \mn@doi [\nat] {10.1038/26421}, \href
  {http://adsabs.harvard.edu/abs/1998Natur.395..355R} {395, 355}

\bibitem[\protect\citeauthoryear{{Rodr{\'{\i}}guez}, {Delgado-Arellano},
  {G{\'o}mez}, {Reipurth}, {Torrelles}, {Noriega-Crespo}, {Raga}  \&
  {Cant{\'o}}}{{Rodr{\'{\i}}guez} et~al.}{2000}]{2000AJ....119..882R}
{Rodr{\'{\i}}guez} L.~F.,  {Delgado-Arellano} V.~G.,  {G{\'o}mez} Y.,
  {Reipurth} B.,  {Torrelles} J.~M.,  {Noriega-Crespo} A.,  {Raga} A.~C.,
  {Cant{\'o}} J.,  2000, \mn@doi [\aj] {10.1086/301231}, \href
  {http://adsabs.harvard.edu/abs/2000AJ....119..882R} {119, 882}

\bibitem[\protect\citeauthoryear{{Rodr{\'{\i}}guez}, {Porras}, {Claussen},
  {Curiel}, {Wilner}  \& {Ho}}{{Rodr{\'{\i}}guez}
  et~al.}{2003}]{2003ApJ...586L.137R}
{Rodr{\'{\i}}guez} L.~F.,  {Porras} A.,  {Claussen} M.~J.,  {Curiel} S.,
  {Wilner} D.~J.,   {Ho} P.~T.~P.,  2003, \mn@doi [\apjl] {10.1086/374882},
  \href {http://adsabs.harvard.edu/abs/2003ApJ...586L.137R} {586, L137}

\bibitem[\protect\citeauthoryear{{Rodr{\'{\i}}guez}, {Gonz{\'a}lez}, {Raga},
  {Cant{\'o}}, {Riera}, {Loinard}, {Dzib}  \& {Zapata}}{{Rodr{\'{\i}}guez}
  et~al.}{2012}]{2012AA...537A.123R}
{Rodr{\'{\i}}guez} L.~F.,  {Gonz{\'a}lez} R.~F.,  {Raga} A.~C.,  {Cant{\'o}}
  J.,  {Riera} A.,  {Loinard} L.,  {Dzib} S.~A.,   {Zapata} L.~A.,  2012,
  \mn@doi [\aap] {10.1051/0004-6361/201117991}, \href
  {http://adsabs.harvard.edu/abs/2012A%26A...537A.123R} {537, A123}

\bibitem[\protect\citeauthoryear{{Sadavoy} et~al.,}{{Sadavoy}
  et~al.}{2010}]{2010ApJ...710.1247S}
{Sadavoy} S.~I.,  et~al., 2010, \mn@doi [\apj] {10.1088/0004-637X/710/2/1247},
  \href {http://adsabs.harvard.edu/abs/2010ApJ...710.1247S} {710, 1247}

\bibitem[\protect\citeauthoryear{{Sargent} \& {Beckwith}}{{Sargent} \&
  {Beckwith}}{1989}]{1989LNP...350..215S}
{Sargent} A.~I.,  {Beckwith} S.~V.~W.,  1989, in {Tenorio-Tagle} G.,  {Moles}
  M.,   {Melnick} J.,  eds,  Lecture Notes in Physics, Berlin Springer Verlag
  Vol. 350, IAU Colloq. 120: Structure and Dynamics of the Interstellar Medium.
  p.~215, \mn@doi{10.1007/BFb0114869}

\bibitem[\protect\citeauthoryear{{Sargent} \& {Beckwith}}{{Sargent} \&
  {Beckwith}}{1994}]{1994ASPC...59..203S}
{Sargent} A.,  {Beckwith} S.,  1994, in {Ishiguro} M.,  {Welch} J.,  eds,
  Astronomical Society of the Pacific Conference Series Vol. 59, IAU Colloq.
  140: Astronomy with Millimeter and Submillimeter Wave Interferometry. p.~203

\bibitem[\protect\citeauthoryear{{Scaife}}{{Scaife}}{2013a}]{2013MNRAS.435.1139S}
{Scaife} A.~M.~M.,  2013a, \mn@doi [\mnras] {10.1093/mnras/stt1361}, \href
  {http://adsabs.harvard.edu/abs/2013MNRAS.435.1139S} {435, 1139}

\bibitem[\protect\citeauthoryear{{Scaife}}{{Scaife}}{2013b}]{2013AdAst2013E..14S}
{Scaife} A.~M.~M.,  2013b, \mn@doi [Advances in Astronomy]
  {10.1155/2013/390287}, \href
  {http://adsabs.harvard.edu/abs/2013AdAst2013E..14S} {2013}

\bibitem[\protect\citeauthoryear{{Scaife} et~al.,}{{Scaife}
  et~al.}{2012}]{2012MNRAS.420.3334S}
{Scaife} A.~M.~M.,  et~al., 2012, \mn@doi [\mnras]
  {10.1111/j.1365-2966.2011.20254.x}, \href
  {http://adsabs.harvard.edu/abs/2012MNRAS.420.3334S} {420, 3334}

\bibitem[\protect\citeauthoryear{{Schwartz} \& {Spencer}}{{Schwartz} \&
  {Spencer}}{1977}]{1977MNRAS.180..297S}
{Schwartz} P.~R.,  {Spencer} J.~H.,  1977, \mnras, \href
  {http://adsabs.harvard.edu/abs/1977MNRAS.180..297S} {180, 297}

\bibitem[\protect\citeauthoryear{{Schwartz}, {Simon}, {Zuckerman}  \&
  {Howell}}{{Schwartz} et~al.}{1984}]{1984ApJ...280L..23S}
{Schwartz} P.~R.,  {Simon} T.,  {Zuckerman} B.,   {Howell} R.~R.,  1984,
  \mn@doi [\apjl] {10.1086/184261}, \href
  {http://adsabs.harvard.edu/abs/1984ApJ...280L..23S} {280, L23}

\bibitem[\protect\citeauthoryear{{Schwartz}, {Simon}  \& {Campbell}}{{Schwartz}
  et~al.}{1986}]{1986ApJ...303..233S}
{Schwartz} P.~R.,  {Simon} T.,   {Campbell} R.,  1986, \mn@doi [\apj]
  {10.1086/164069}, \href {http://adsabs.harvard.edu/abs/1986ApJ...303..233S}
  {303, 233}

\bibitem[\protect\citeauthoryear{{Skinner} \& {Brown}}{{Skinner} \&
  {Brown}}{1994}]{1994AJ....107.1461S}
{Skinner} S.~L.,  {Brown} A.,  1994, \mn@doi [\aj] {10.1086/116959}, \href
  {http://adsabs.harvard.edu/abs/1994AJ....107.1461S} {107, 1461}

\bibitem[\protect\citeauthoryear{{Snell}, {Loren}  \& {Plambeck}}{{Snell}
  et~al.}{1980}]{1980ApJ...239L..17S}
{Snell} R.~L.,  {Loren} R.~B.,   {Plambeck} R.~L.,  1980, \mn@doi [\apjl]
  {10.1086/183283}, \href {http://adsabs.harvard.edu/abs/1980ApJ...239L..17S}
  {239, L17}

\bibitem[\protect\citeauthoryear{{Spencer} \& {Schwartz}}{{Spencer} \&
  {Schwartz}}{1974}]{1974ApJ...188L.105S}
{Spencer} J.~H.,  {Schwartz} P.~R.,  1974, \mn@doi [\apjl] {10.1086/181445},
  \href {http://adsabs.harvard.edu/abs/1974ApJ...188L.105S} {188, L105}

\bibitem[\protect\citeauthoryear{{Stapelfeldt} et~al.,}{{Stapelfeldt}
  et~al.}{1998}]{1998ApJ...508..736S}
{Stapelfeldt} K.~R.,  et~al., 1998, \mn@doi [\apj] {10.1086/306422}, \href
  {http://adsabs.harvard.edu/abs/1998ApJ...508..736S} {508, 736}

\bibitem[\protect\citeauthoryear{{Stojimirovi{\'c}}, {Narayanan}, {Snell}  \&
  {Bally}}{{Stojimirovi{\'c}} et~al.}{2006}]{2006ApJ...649..280S}
{Stojimirovi{\'c}} I.,  {Narayanan} G.,  {Snell} R.~L.,   {Bally} J.,  2006,
  \mn@doi [\apj] {10.1086/506340}, \href
  {http://adsabs.harvard.edu/abs/2006ApJ...649..280S} {649, 280}

\bibitem[\protect\citeauthoryear{{Strom}, {Strom}  \& {Vrba}}{{Strom}
  et~al.}{1976}]{1976AJ.....81..320S}
{Strom} K.~M.,  {Strom} S.~E.,   {Vrba} F.~J.,  1976, \mn@doi [\aj]
  {10.1086/111891}, \href {http://adsabs.harvard.edu/abs/1976AJ.....81..320S}
  {81, 320}

\bibitem[\protect\citeauthoryear{{Torres}, {Loinard}, {Mioduszewski}, {Boden},
  {Franco-Hern{\'a}ndez}, {Vlemmings}  \& {Rodr{\'{\i}}guez}}{{Torres}
  et~al.}{2012}]{2012ApJ...747...18T}
{Torres} R.~M.,  {Loinard} L.,  {Mioduszewski} A.~J.,  {Boden} A.~F.,
  {Franco-Hern{\'a}ndez} R.,  {Vlemmings} W.~H.~T.,   {Rodr{\'{\i}}guez} L.~F.,
   2012, \mn@doi [\apj] {10.1088/0004-637X/747/1/18}, \href
  {http://adsabs.harvard.edu/abs/2012ApJ...747...18T} {747, 18}

\bibitem[\protect\citeauthoryear{{Weintraub}, {Sandell}  \&
  {Duncan}}{{Weintraub} et~al.}{1989a}]{1989ApJ...340L..69W}
{Weintraub} D.~A.,  {Sandell} G.,   {Duncan} W.~D.,  1989a, \mn@doi [\apjl]
  {10.1086/185441}, \href {http://adsabs.harvard.edu/abs/1989ApJ...340L..69W}
  {340, L69}

\bibitem[\protect\citeauthoryear{{Weintraub}, {Zuckerman}  \&
  {Masson}}{{Weintraub} et~al.}{1989b}]{1989ApJ...344..915W}
{Weintraub} D.~A.,  {Zuckerman} B.,   {Masson} C.~R.,  1989b, \mn@doi [\apj]
  {10.1086/167859}, \href {http://adsabs.harvard.edu/abs/1989ApJ...344..915W}
  {344, 915}

\bibitem[\protect\citeauthoryear{{White}, {Bicknell}, {McGregor}  \&
  {Salmeron}}{{White} et~al.}{2014}]{2014MNRAS.442...28W}
{White} M.~C.,  {Bicknell} G.~V.,  {McGregor} P.~J.,   {Salmeron} R.,  2014,
  \mn@doi [\mnras] {10.1093/mnras/stu788}, \href
  {http://cdsads.u-strasbg.fr/abs/2014MNRAS.442...28W} {442, 28}

\bibitem[\protect\citeauthoryear{{Woody}, {Scott}, {Scoville}, {Mundy},
  {Sargent}, {Padin}, {Tinney}  \& {Wilson}}{{Woody}
  et~al.}{1989}]{1989ApJ...337L..41W}
{Woody} D.~P.,  {Scott} S.~L.,  {Scoville} N.~Z.,  {Mundy} L.~G.,  {Sargent}
  A.~I.,  {Padin} S.,  {Tinney} C.~G.,   {Wilson} C.~D.,  1989, \mn@doi [\apjl]
  {10.1086/185374}, \href {http://adsabs.harvard.edu/abs/1989ApJ...337L..41W}
  {337, L41}

\bibitem[\protect\citeauthoryear{{Wu}, {Takakuwa}  \& {Lim}}{{Wu}
  et~al.}{2009}]{2009ApJ...698..184W}
{Wu} P.-F.,  {Takakuwa} S.,   {Lim} J.,  2009, \mn@doi [\apj]
  {10.1088/0004-637X/698/1/184}, \href
  {http://adsabs.harvard.edu/abs/2009ApJ...698..184W} {698, 184}

\makeatother
\end{thebibliography}



\appendix

\section{SED Reference Tables}
\label{appA}

An extensive literature search was conducted for integrated flux densities (on scales comparable with the GMRT) to include in the spectral energy distributions following \citet{2012MNRAS.423.1089A}. Where uncertainties were not provided, an error of 10\,per~cent was used in the model fittings and this is indicated by a $^{\dag}$. The reference list for the flux densities used in the L1551~IRS~5 spectral energy distribution (for $\nu<1$\,THz and in addition to the data presented in this work), can be found in the online supplementary material of \citet{2012MNRAS.423.1089A}.

\begin{table}
\centering
\caption{Reference list for the spectral energy distribution data for
T~Tau. Where uncertainties were not provided, an error of 10\,per~cent
was used in the model fittings and this is indicated by a ${\dag}$.}
\label{tab:sedrefTTAU}
\begin{tabular}{d{-1}@{}d{-1}@{}c@{~~}d{-1}l}\hline\hline
\multic{1}{$\nu$ (GHz)} & \multic{3}{$S_{\nu}$ (mJy)} & Reference \\\hline
 0.323  &     3.95 & $\pm$ & 0.29  & this work \\
 0.608  &     4.24 & $\pm$ & 0.23  & this work \\
 1.4    &     4.80 &    & \dagmark & \citet{1998AJ....115.1693C}\\
 1.4    &     2.80 & $\pm$ & 0.50  & \citet{1986ApJ...303..233S}\\
 1.465  &     5.00 & $\pm$ & 1.00  & \citet{1983RMxAA...8..163R}\\
 1.465  &     3.00 & $\pm$ & 1.00  & \citet{1984ApJ...280L..23S}\\
 1.49   &     2.74 & $\pm$ & 0.11  & \citet{1986AJ.....92.1396C}\\
 2.695  &     4.00 & $\pm$ & 2.00  & \citet{1974ApJ...188L.105S}\\
 4.86   &     4.94 & $\pm$ & 0.34  & \citet{1994AJ....107.1461S}\\
 4.86   &     4.62 & $\pm$ & 0.26  & \citet{1994AJ....107.1461S}\\
 4.86   &     4.53 & $\pm$ & 0.29  & \citet{1994AJ....107.1461S}\\
 4.885  &     5.80 & $\pm$ & 0.60  & \citet{1982ApJ...253..707C}\\
 4.885  &     5.00 & $\pm$ & 0.60  & \citet{1983RMxAA...8..163R}\\
 4.885  &     4.30 & $\pm$ & 0.20  & \citet{1984ApJ...280L..23S}\\
 4.885  &     5.20 & $\pm$ & 0.50  & \citet{1984ApJ...282..699B}\\
 4.885  &     5.70 & $\pm$ & 0.50  & \citet{1986ApJ...303..233S}\\
 4.885  &     6.40 &    & \dagmark & \citet{1987ApJ...320..364E}\\
 5      &     6.50 & $\pm$ & 0.07  & \citet{1982ApJ...253..707C}\\
 8.085  &     8.00 & $\pm$ & 2.00  & \citet{1974ApJ...188L.105S}\\
 8.44   &     5.19 & $\pm$ & 0.30  & \citet{1994AJ....107.1461S}\\
 8.44   &     5.05 & $\pm$ & 0.31  & \citet{1994AJ....107.1461S}\\
 8.44   &     5.79 & $\pm$ & 0.34  & \citet{1994AJ....107.1461S}\\
 8.44   &     7.93 & $\pm$ & 0.51  & \citet{1994AJ....107.1461S}\\
 8.44   &     5.81 & $\pm$ & 0.41  & \citet{1994AJ....107.1461S}\\
 10.5   &     7.30 & $\pm$ & 2.80  & \citet{1979AJ.....84.1709M}\\
 10.69  &     8.00 & $\pm$ & 2.00  & \citet{1976AA....46...11A}\\
 14.94  &     6.50 & $\pm$ & 0.07  & \citet{1986AJ.....92.1396C}\\
 14.949 &     5.14 & $\pm$ & 0.43  & \citet{1994AJ....107.1461S}\\
 14.965 &     9.00 & $\pm$ & 1.00  & \citet{1984ApJ...280L..23S}\\
 14.965 &     7.50 & $\pm$ & 0.70  & \citet{1984ApJ...282..699B}\\
 14.965 &    11.10 & $\pm$ & 1.00  & \citet{1986ApJ...303..233S}\\
 15     &     8.60 & $\pm$ & 2.00  & \citet{1982AA...107..368B}\\
 86     &     32.7 & $\pm$ & 7.8   & \citet{1986AA...164..227A}\\
 90     &     68.0 & $\pm$ & 27.0  & \citet{1977MNRAS.180..297S}\\
 112.6  &     75.0 & $\pm$ & 7.5   & \citet{1989ApJ...344..915W}\\
 226    &     894  & $\pm$ & 153   & \citet{1986AA...164..227A}\\
 230    &     280  & $\pm$ & 9     & \citet{1990AJ.....99..924B}\\
 230    &     253  & $\pm$ & 18    & \citet{1993AA...273..221R}\\
 231    &     280  & $\pm$ & 9     & \citet{2005ApJ...631.1134A}\\
 250    &     296  & $\pm$ & 25    & \citet{1994AA...281..161A}\\
 272    &     320  & $\pm$ & 30    & \citet{1990ApJ...357..606A}\\
 272    &     579  & $\pm$ & 27    & \citet{1989ApJ...340L..69W}\\
 284    &     417  & $\pm$ & 41    & \citet{1991ApJ...381..250B}\\
 341    &     304  &   & \dagmark  & \citet{2012ApJ...751..115H}\\
 353    &    2250  &   & \dagmark  & \citet{2010ApJ...710.1247S}\\
 353    &    2470  &   & \dagmark  & \citet{2008ApJS..175..277D}\\
 353    &      628 & $\pm$ & 17    & \citet{2005ApJ...631.1134A}\\
 375    &     1216 & $\pm$ & 44    & \citet{1989ApJ...340L..69W}\\
 375    &      860 & $\pm$ & 80    & \citet{1994AJ....108..661J}\\
 380    &     1070 & $\pm$ & 110   & \citet{1990ApJ...357..606A}\\
 390    &      910 & $\pm$ & 90    & \citet{1991ApJ...381..250B}\\
 463    &     1300 & $\pm$ & 200   & \citet{1990ApJ...357..606A}\\
 676    &     2600 & $\pm$ & 400   & \citet{1990ApJ...357..606A}\\
 676    &     1655 & $\pm$ & 218   & \citet{2005ApJ...631.1134A}\\
 676    &     6710 & $\pm$ & 610   & \citet{1989ApJ...340L..69W}\\
 854    &    10170 & $\pm$ & 730   & \citet{1989ApJ...340L..69W}\\
 854    &     8149 & $\pm$ & 253   & \citet{2005ApJ...631.1134A}\\\hline
\end{tabular}
\end{table}

\begin{table}
\centering
\caption{Reference list for the spectral energy distribution data for
DG~Tau. Where uncertainties were not provided, an error of 10\,per~cent
was used in the model fittings and this is indicated by a ${\dag}$.}
\label{tab:sedrefDGTAU}
\begin{tabular}{d{-1}@{}d{-1}@{}c@{~~}d{-1}l}\hline\hline
\multic{1}{$\nu$ (GHz)} & \multic{3}{$S_{\nu}$ (mJy)} & Reference \\\hline
 0.608  &  0.58 & $\pm$ & 0.20 & this work\\
 3.0    &  0.76 & $\pm$ & 0.04 & Ainsworth et~al., in prep.\\
 4.89   &  0.70 & $\pm$ & 0.10 & \citet{1982ApJ...253..707C} \\
 4.89   &  1.20 & $\pm$ & 0.10 & \citet{1984ApJ...282..699B} \\
 4.89   &  1.03 & $\pm$ & 0.07 & \citet{1986AJ.....92.1396C} \\
 4.89   &  0.60 & $\pm$ & 0.10 & \citet{1987ApJ...320..364E} \\
 5.4    &  1.05 & $\pm$ & 0.05 & \citet{2013ApJ...766...53L} \\
 8.5    &  1.27 & $\pm$ & 0.05 & \citet{2012AA...537A.123R} \\
 8.5    &  1.29 & $\pm$ & 0.07 & \citet{2013ApJ...766...53L} \\
 8.5    &  1.10 & $\pm$ & 0.06 & \citet{2013ApJ...766...53L} \\
 15     &  2.01 &  & \dagmark  & \citet{1986AJ.....92.1396C} \\
 16.12  &  1.40 & $\pm$ & 0.07 & \citet{2012MNRAS.420.3334S} \\
 23     &  2.90 &  & \dagmark  & \citet{1982AA...107..368B} \\
 43.3   &  3.90 & $\pm$ & 0.22 & \citet{2013ApJ...766...53L} \\
 98     &  57.0 & $\pm$ & 4.0  & \citet{1991AJ....102.2054O} \\
 110    &  71 &  & \dagmark    & \citet{1989LNP...350..215S} \\
 110    &  72 &  & \dagmark    & \citet{1996ApJ...457..277K} \\
 111    &  75 &  & \dagmark    & \citet{1994ASPC...59..203S} \\
 112    &  80 &  & \dagmark    & \citet{1989ApJ...337L..41W} \\
 147    &  186 & $\pm$ & 17    & \citet{1996ApJ...465L.137K} \\
 230    &  443 &  & \dagmark   & \citet{1990AJ.....99..924B} \\
 230    &  440 &  & \dagmark   & \citet{1990ApJ...357..606A} \\
 250    &  420 & $\pm$ & 42    & \citet{1994AA...281..161A} \\
 260    &  489 &  & \dagmark   & \citet{1991ApJ...381..250B} \\
 270    &  532 &  & \dagmark   & \citet{1989ApJ...340L..69W} \\
 273    &  440 & $\pm$ & 30    & \citet{1990ApJ...357..606A} \\
 273    &  532 & $\pm$ & 48    & \citet{1989ApJ...340L..69W} \\
 284    &  489 & $\pm$ & 33    & \citet{1991ApJ...381..250B} \\
 375    &  940 & $\pm$ & 100   & \citet{1990ApJ...357..606A} \\
 390    &  860 & $\pm$ & 100   & \citet{1991ApJ...381..250B} \\
 480    &  1210 & $\pm$ & 200  & \citet{1991ApJ...381..250B} \\
 500    &  1400 & $\pm$ & 200  & \citet{1990ApJ...357..606A} \\
 666    &  3390 & $\pm$ & 440  & \citet{1989ApJ...340L..69W} \\
 666    &  3600 & $\pm$ & 400  & \citet{1990ApJ...357..606A} \\
 857    &  9600 & $\pm$ & 1100 & \citet{1990ApJ...357..606A} \\\hline
\end{tabular}
\end{table}


\bsp	
\label{lastpage}
\end{document}